\documentclass[11pt,eqsecnum]{revtex4}

\usepackage{bm}  
\usepackage{graphicx}
\usepackage{amssymb}
\usepackage{color}
\usepackage{amscd}
\usepackage{epsfig}

\newcommand{\bfr}{{\bf r}}

\newcommand{\bfk}{{\bf k}}

\newcommand{\bfT}{{\bf T}}
\newcommand{\bnabla}{{\bm \nabla}}
\newcommand{\bxi}{{\bm \xi}}

\newcommand{\talpha}{{\tilde{\alpha}}}

\newcommand{\tm}{{\tilde{m}}}

\newcommand{\nhat}{\hat{u}}
\newcommand{\nvec}{{\hat{\bm{u}}}}
\newcommand{\phat}{\hat{p}}
\newcommand{\pvec}{{\hat{\bf p}}}

\newcommand{\xhat}{{\bf \hat{x}}}

\begin{document}
\title{Hydrodynamics of isotropic and liquid crystalline  active polymer solutions}
\author{Aphrodite Ahmadi}
\affiliation{Physics Department, Syracuse University, Syracuse, NY
13244, USA}
\author{T. B. Liverpool}
\affiliation{Department of Applied Mathematics, University of
Leeds, Woodhouse Lane, Leeds LS2 9JT, UK}
\author{M. C. Marchetti}
\affiliation{Physics Department, Syracuse University, Syracuse, NY
13244, USA}

\date{\today}

\begin{abstract}
  We describe the large-scale collective behavior of solutions of
  polar biofilaments and both stationary and mobile crosslinkers.
  Both mobile and stationary crosslinkers induce filament alignment
  promoting either polar or nematic order. In addition, mobile
  crosslinkers, such as clusters of motor proteins, exchange forces
  and torques among the filaments and render the homogeneous states
  unstable via filament bundling.  We start from a Smoluchowski
  equation for rigid filaments in solutions, where pairwise
  crosslink-mediated interactions among the filaments yield
  translational and rotational currents. The large-scale properties of
  the system are described in terms of continuum equations for
  filament and motor densities, polarization and alignment tensor
  obtained by coarse-graining the Smoluchovski equation. The possible
  homogeneous and inhomogeneous states of the systems are obtained as
  stable solutions of the dynamical equations and are characterized in
  terms of experimentally accessible parameters.  We make contact with
  work by other authors and show that our model allows for an estimate
  of the various parameters in the hydrodynamic equations in terms of
  physical properties of the crosslinkers.
\end{abstract}
\pacs{87.16.-b,47.54.+r,05.65.+b}
\maketitle

\section{Introduction}
\label{sec:Introduction}

Soft active systems are a new and exciting class of complex fluids
to which energy is continuously being supplied by internal or
external sources. Biology provides many examples of such systems
which include cell membranes, biopolymer solutions driven by
chemical reactions, collections of living cells moving on a
substrate, and the cytoskeleton of eukariotic
cells~\cite{Alberts}. The cytoskeleton is a complex
three-dimensional network of long filamentary proteins (mainly
F-actin and microtubules) cross-linked by  a variety of smaller
proteins~\cite{howard,dogterom95}. Among the latter are clusters
of \emph{active} motor proteins, such as myosin and kinesin,  that
transform chemical energy from the hydrolysis of ATP (adenosine tri-phosphate) 
into mechanical work and are capable of "walking" along the filaments,
mediating the exchange of forces between
them~\cite{takiguchi,urrutia,nedelec97,surrey01}.

The self-organization of motor-filament mixtures has been the
subject of recent experiments~\cite{takiguchi,urrutia,nedelec97,surrey01}.
Specifically, mixtures of microtubules and associated motor
clusters have been studied in vitro in a confined
quasi-two-dimensional geometry~\cite{nedelec97,surrey01}. Complex
patterns, including asters and vortices or spirals have been
observed in these in-vitro experiments as a function of motor and
ATP concentration~\cite{nedelec97,surrey01}. The high frequency
mechanical response of  active filament solutions which are
dominated by the bending modes of the filaments have also been
studied both experimentally and
theoretically~\cite{humphrey02,TBL_active,TBL_active2}. The study of the
properties of these simplified model systems paves the way to a
better understanding of  the formation and stability of  more
complex structures of biological relevance, such as the mitotic
spindle formed during cell division~\cite{Alberts}.



There have been a number of recent theoretical studies of the
collective dynamics of rigid active filaments.  First and most
microscopic, numerical simulations with detailed modeling of the
filament-motor coupling have been used to generate patterns
similar to those found in experiments~\cite{nedelec97,surrey01}.
These approaches have given valuable insights into the problem but
are limited to small system sizes by computing power.  A second
very interesting development has been the proposal of 'mesoscopic'
mean-field kinetic equations governing the dynamics of individual
filaments where the effect of motors was incorporated via a
motor-induced relative velocity of pairs of filaments, with the
form of such velocity inferred from general symmetry
considerations~\cite{nakazawa96,SNbook,Kruse00,Kruse03,Kruse01}.  Finally,
hydrodynamic equations have also been proposed where the large
scale dynamics of the mixture is described in terms of a few
coarse-grained fields whose dynamics is also inferred from
symmetry considerations, 
~\cite{bassetti00,Lee01,ramaswamy02,Kim03,ramaswamy04,Kruse04,Kruse05,Sankararaman03}.
Recently, a connection between the mesoscopic and hydrodynamic
approaches was established by us by deriving hydrodynamical
equations via a coarse-graining of the kinetic
equations~\cite{TBLMCM03}. This was done in the spirit of polymer
physics which has been successful at predicting macroscopic
dynamical behavior of polymer solutions based on models of the
microscopic dynamics. To make a link with the motor properties we
consider a simplified model of the motor filament interaction in
Appendix \ref{app:microscopic}.

The richness of the phenomena exhibited by the cytoskeleton is
illustrated by the ability of its constituents to organize in a variety of
different structures. In addition different constituents can form
very similar structures. This leads naturally to the question -
how much of the behavior is specific and how much is generic? To
answer this question it is important to make the connection
between microscopic models and 'generic' hydrodynamic approaches.

In this paper we describe a derivation of the hydrodynamic equation
for a solution of polar filaments and both stationary and mobile
crosslinkers. A brief summary of the approach and some of the results
have been presented earlier~\cite{TBLMCM03,AATBLMCM05}. The filaments
are modeled as rigid rods of fixed length. Hydrodynamics is obtained
by coarse-graining the Smoluchowski equation for rods in solution,
coupled via excluded volume and motor-mediated interactions. Small
protein clusters crosslinking the filaments can be grouped in two
classes. The first class comprises stationary crosslinkers, such as
$\alpha$-actinin, that can induce rotation and alignment of the
filaments even in the limit of vanishing ATP consumption. Such passive
crosslinkers may be polar or nonpolar in nature depending on whether
they preferentially bind to pairs of filaments of the same polarity or
their binding rate is independent of the filaments' polarity.  They
always induce filament alignment via a mechanism that has been
referred to as "zipping" effect in the literature \cite{Uhde2004}. In
general we expect that most crosslinkers will be polar, although
"disordered" motor clusters (i.e, cluster with no spatial order in
the arrangements of individual motors as in e.g. {\em small} myosin clusters) can crosslink filaments
regardless of their relative polarity.  Stationary crosslinkers can
lead to the onset of the homogeneous nematic and polarized states. The
interplay between these two types of order is determined by the
crosslinkers' polarity. The second class consists of clusters of motor
proteins crosslinking two filaments, ``active crosslinkers''.  These
can also drive the system into nematic and polarized states.  However,
in addition by consuming ATP, the motor heads can "walk" along the
filaments and mediate the exchange of forces between filaments,
inducing filament motion relative to the solution (treated here as an
inert background).  The motor activity depends crucially on the ATP
consumption rate, which is the driving force that sets up and
maintains the nonequilibrium state and enters the equation as a
chemical potential.  Motor activity destabilizes the homogeneous
states and induces the formation of spatially inhomogeneous structures
on mesoscopic scales, reminiscent of those seen in the in vitro
experiments. There are two main motor-mediated mechanisms for force
exchange among the filaments.  First, active crosslinkers induce
bundling of filaments 
, building up density inhomogeneities. This is the main mechanism
responsible for instabilities. It is effective only if the rate at
which motor clusters step along the filament is inhomogeneous, which
can be due to crowding and fluctuations in the density of bound
motors, or to stalling at the polar end. In addition, active
crosslinkers sort the filaments according to polarization at a rate
proportional to the mean motor stepping rate. This mechanism is
important in the polarized state, where it yields filament advection
along the direction of polarization and allows for the onset of
oscillatory structures.

The forces and torques exchanged by filaments via the crosslinks
are described by considering the kinematics of two filaments
crosslinked by a single protein cluster that can rotate and
translate as a rigid object relative to the filaments. The
hydrodynamic equations are then obtained by suitable
coarse-graining of the Smoluchowski equation. This method yields a
general form of hydrodynamics which incorporates all terms allowed
by symmetry, yet it provides a connection between the
coarse-grained and the microscopic dynamics. By comparing the
equations obtained here to those obtained from a microscopic model
of the forces exchanged between motors and filaments we can relate
some of the parameters in the hydrodynamic equations to parameters
that can be controlled in experiments.

The hydrodynamic equations are then used to describe
the dynamics of the isotropic, nematic and polarized solutions. We
characterize the possible homogeneous states of the system in term
of experimentally accessible parameters and discuss the various
mechanisms by which motor activity can destabilize each
homogeneous state.

In Section \ref{sec:Model} we describe the kinetic model of rods
crosslinked by small protein clusters and set up the formalism of
the Smoluchwski equation. The dependence of the crosslinked-induced
rotational and translational velocities of the filaments
on filament orientation and position is obtained from general
symmetry considerations and conservation laws. The details of the
kinematics of motors and filaments are described in Appendix
\ref{app:microscopic}, where a specific microscopic model of the
coupling is also presented. In Section \ref{sec:Hydrodynamic
equations} we obtain the hydrodynamic equations for the system by
a systematic coarse graining of the Smoluchowski equation. The
full form of the hydrodynamic equations, including diffusive,
excluded volume and active contributions, is given in Appendix
\ref{app:currents}. The nonlinear hydrodynamic equations are
solved in Section \ref{sec:Homogeneous_states} to obtain the
possible homogeneous steady states of the system. "Phase diagrams"
are constructed in terms of the filament and crosslinkers
densities identifying the isotropic, nematic and polarized
states. The nonlinear hydrodynamic equations for each homogeneous
state are presented in Section \ref{sec:inhomogeneous}, where the
stability of each state is also studied. All homogeneous states
become unstable at high filament and crosslinkers densities via
filament bundling. The interplay of bundling and diffusion
promotes the onset of stable spatial
structures on mesoscopic scales. Finally, we conclude with a
discussion of open questions and a comparison with related work.

\section{The model: Smoluchowski equation for motor-filaments solutions}
\label{sec:Model} We model the system as a collection of thin rods
of fixed length $l$ and diameter $b<<l$ crosslinked by small protein
clusters (of linear size $\sim b$) that can exchange torques and
forces between the filaments. Filaments and crosslinkers move through
a solvent which is assumed inert. The solution forms a
quasi-two-dimensional film, of thickness much smaller than the length
of the filaments. The dynamics of both filaments and crosslinkers is
overdamped. This is a good model for a quiescent solution with no
externally imposed flow nor net flow generated by motor activity. We
are interested in describing the filament dynamics on time scales
large compared to the characteristic times for binding and unbinding
of the crosslinkers so that we can treat a constant fraction of them
as bound.  The dynamics of crosslinkers binding and unbinding was
considered for instance in Ref.~\cite{Sankararaman03} and 
it was found that varying the rates of binding and unbinding of motor
clusters
did not affect the nature of the nonequilibrium steady states of the
active solution. The temperature of the system is taken to be constant
and the effect of thermal fluctuations is not considered explicitly.
We assume, however, that the stochastic nature of the crosslinkers
dynamics, as well as other sources of noise in the systems, can be
incorporated in an effective temperature $T_a$ that may differ from
the actual temperature of the solution \cite{TBL_active,TBL_active2}.
Finally, although the kinetic model described below applies to a
solution with a low concentration of filaments, the structure of the
continuum equations obtained upon coarse-graining the kinetic model is
general and not restricted to low density. On the other hand, the
quantitative estimates obtained for the various parameters in the
hydrodynamic equations are for a low density of filaments and
crosslinkers.

\begin{figure}
\center \resizebox{0.49\textwidth}{!}{
  \includegraphics{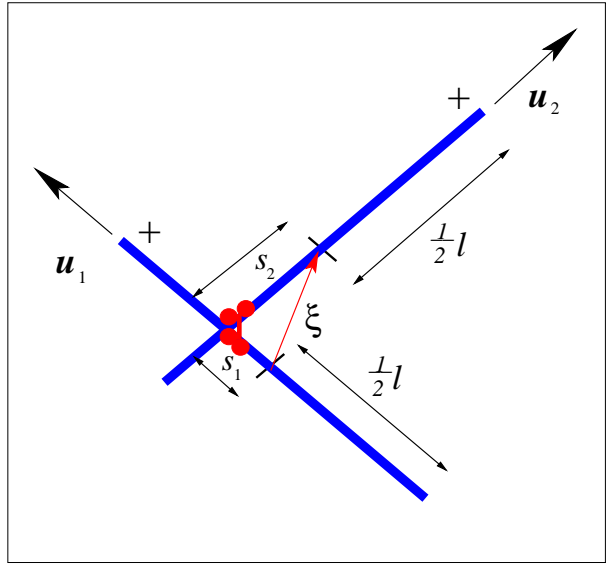}}

\caption{The geometry of overlap between two interacting filaments of length $l$ cross-linked by an active cluster. The cross-link is a distance $s_1,(s_2)$ from the centre of mass of filament $1(2)$. The distance between centres $\bm{\xi}= {\bf r}_2-{\bf r}_1 = s_1 \nvec_1 - s_2 \nvec_2$.}
\label{2filam_coord}       
\end{figure}

The dynamics of the concentration $c({\bf r},\nvec,t)$ of filaments
with center of mass at ${\bf r}$ and orientation $\nvec$ at time $t$
is governed by the Smoluchowski equation ~\cite{DoiEdwards,Shimada88},
which describes conservation of the number of filaments,
\begin{equation}
\partial_t c=-\bm{\nabla}\cdot{\bf J}_c - \bm{\mathcal R} \cdot \bm{\mathcal
J}_c\;,
\end{equation}
where  $\bm{\mathcal R}=\nvec\times\partial_{\nvec}$ is the
rotation operator. The {\em translational} current density, ${\bf
J}_c$, and {\em rotational} current density, $\bm{\mathcal J}_c$,
are given by
\begin{eqnarray}
\label{current}
J_{ci}&=&-D_{ij}\nabla_{j}c-\frac{D_{ij}}{k_BT_a}c~\nabla_{j}V_{\rm
ex}+J_{ci}^{\rm A}\;, \\
\label{current2}{\mathcal J}_{ci} &=& -D_r {\cal R}_i c - {D_r \over k_B T_a} c
{\mathcal{R}}_iV_{\rm ex}+\mathcal{J}_{ci}^{\rm A}\;,
\end{eqnarray}
where
$D_{ij}=D_\parallel\nhat_i\nhat_j+D_\perp\big(\delta_{ij}-\nhat_i\nhat_j\big)$
is the translational diffusion tensor and $D_r$ is the rotational
diffusion rate. For a low-density solution of long, thin rods
$D_\perp=D_\parallel/2\equiv D/2$, where $D=k_BT_a\ln(l/b)/(2\pi\eta
l)$, with $\eta$ the solvent viscosity, and $D_r=6D/l^2$. The
potential $V_{\rm ex}$ incorporates excluded volume effects which give
rise to the nematic transition in a solution of hard rods. It can be
written by generalizing the Onsager interaction to inhomogeneous
systems as $k_BT_a$ times the probability of finding another rod
within the interaction area of a given rod (see Figure
\ref{2filam_coord}). In two dimensions this gives
\begin{eqnarray}
\label{Vex} V_{\rm ex}(\bfr_1,\nvec_1)&=&k_BT_a\int d\bfr_2\int
d\nvec_2~c(\bfr_2,\nvec_2,t)~|\nvec_1\times\nvec_2|\int_{s_1 s_2}
\delta(\bfr_1+\nvec_1s_1-\bfr_2-\nvec_2s_2)\nonumber\\
&=& k_BT_a\int
d\nvec_2\int_{s_1 s_2}|\nvec_1\times\nvec_2|~c({\bf
r}_1+\bm{\xi},\nvec_2,t)\;,
\end{eqnarray}
where $s_i$, with $-l/2\leq s_i\leq l/2$, parametrizes the
position along the length of the $i$-th filament, for $i=1,2$, and
$\int_{s_i}...\equiv \int_{-l/2}^{l/2}~ds_i...$. The
$\delta$-function ensures that the filaments be within each
other's interaction volume, i.e., in the thin rod limit $b<<l$
considered here, have a point of contact. The factor
$|\nvec_1\times\nvec_2|$ represents the excluded area of two thin
filaments of orientation $\nvec_1$ and $\nvec_2$ touching at one
point. In the second equality we let $\bm{\xi}={\bf r}_2-{\bf
r}_1 = \nvec_1s_1-\nvec_2s_2$~\cite{DoiEdwards}.

The translational and rotational active current of filaments with
center of mass at ${\bf r}_1$ and orientation along $\nvec_1$ are
written as

\begin{eqnarray}
&&{\bf J}_c^{\rm A}({\bf
r}_1,\nvec_1)=b^2\int_{\nvec_2}\int_{s_1 s_2}|\nvec_1\times\nvec_2|m({\bf
r}_1+\nvec_1s_1){\bf
v}_1(s_1,s_2,\nvec_1,\nvec_2) c({\bf
r}_1,\nvec_1,t) c({\bf r}_1+\bm\xi,\nvec_2,t)\label{Jtact}\\
&&\bm{\mathcal J}_c^{\rm A}({\bf r}_1,\nvec_1)=b^2\int
_{\nvec_2}\int_{s_1 s_2}|\nvec_1\times\nvec_2|
m({\bf
r}_1+\nvec_1s_1)\bm\omega_1(s_1,s_2,\nvec_1,\nvec_2)
c({\bf r}_1,\nvec_1,t) c({\bf r}_1+\bm\xi,\nvec_2,t)\label{Jract}\;.
\end{eqnarray}
where $m({\bf r})$ is the density of bound crosslinkers, evaluated
at the point of attachment to the filaments. Also, ${\bf
v}_1=\dot{\bf r}_1$ is the translational velocity that the center
of mass of filament 1 acquires due to the crosslinker-mediated
interaction with filament 2, when the centers of mass of the two
filaments are separated by $\bm\xi$. Similarly,
$\bm\omega_1 \times \nvec_1 =\dot{\nvec}_1$ 
is the crosslinker-induced velocity of rotation about the center of
mass (see Figure \ref{2filam_coord}).

Small clusters of motor proteins crosslinking two filaments can be
stationary or mobile (active). In general the density of bound crosslinkers
can be written as $m=m_s+m_a$, with $m_s$ and $m_a$ the
densities of stationary and mobile crosslinks, respectively.
Mobile crosslinks are clusters of motor proteins that can diffuse
and be convected along the filaments. The mobile crosslinker
density obeys a diffusion-convection equation given by
\begin{equation}
\label{mdiff}
\partial_tm_a=D_m\nabla^2m_a-\bm{\nabla}\cdot{\bf J}_m\;,
\end{equation}
where
\begin{equation}
\label{Jm} {\bf J}_m=\frac{b^2}{l}\int_\nvec\int_s
\nvec~u(s)~c(\bfr,\nvec,t)m_a(\bfr+\nvec s,t)\;,
\end{equation}
and $u(s)$ is the speed at which a motor cluster steps on a
filament at position $s$.  The mean value $u_0=\int_s u(s)/l$ of
the stepping rate is $u_0\sim a R_{ATP}$, where $a$ is
the step size and $R_{ATP}$ is the ATP consumption rate. For
typical motor clusters (kinesins on microtubules or myosins on
F-actin) $u_0\sim nm/msec$~\cite{howard}. As shown in Ref. \cite{TBLMCM05},
spatial inhomogeneities in the motor stepping rate $u(s)$,
especially the stalling of motors at the polar end, are crucial for
driving filament bundling and pattern formation. Such
inhomogeneities may arise from motor crowding or from large
fluctuations in the concentration of ATP under condition of near
depletion. Very recent experiments in purified actin-myosin II
solutions have indeed suggested that the motor-driven formation of
spatially inhomogeneous patterns, such as asters and vortices, may
be associated with strong inhomogeneities in motor activity~\cite{Smith}.

The translational and rotational velocities of the filaments
induced by crosslinkers are written in a general form that is
consistent with translational and rotational invariance. We
consider a pair of filaments cross-linked by a single protein
cluster. As seen below, all crosslinkers can exchange
torque among the filament and induce filament alignment or
"zipping". Mobile crosslinkers that consume ATP to step along
filaments can also exchange forces and induce translational motion
of the filaments. In general the rotational and translational
dynamics induced by the crosslinkers is coupled.

It is convenient to introduce the relative velocity and net
velocity of the filament pair as
\begin{eqnarray}
&&{\bf v}={\bf v}_1-{\bf v}_2\;,\nonumber\\
&&{\bf V}=\frac{{\bf v}_1+{\bf v}_2}{2}\;,
\end{eqnarray}
with ${\bf v}_{1,2}={\bf V}\pm{\bf v}/2$. A general form of the
relative linear velocity ${\bf v}$ and angular velocity
$\bm\omega=\bm\omega_1-\bm\omega_2$ of the filament pair
consistent with symmetries and conservation laws is
\begin{eqnarray}
&& {\bf v}=\frac{\talpha(\theta)}{2l}~\bm\xi
+\frac{\beta(\theta)}{2}(\nvec_2-\nvec_1)\;,\label{vrel}\\
&&
{\bm\omega}=4\gamma(\theta)~\nvec_1\times\nvec_2\label{omega}\;,
\end{eqnarray}
where $\bm\xi=\nvec_1s_1-\nvec_2s_2$ is the separation of the
filaments' centers of mass, and $\talpha$, $\beta$ and $\gamma$
depend on the relative orientation of the two filaments through
the angle $\theta=\cos^{-1}(\nvec_1\cdot\nvec_2)$.  The angular
dependence of $\talpha$, $\beta$ and $\gamma$ arises both from the
kinematics of the crosslinker-mediated filament interaction, as
well as from the dependence of the probability that a protein
cluster binds two filaments on the angle between the
filaments at contact.

It is instructive to rewrite the relative velocity ${\bf v}$ in
terms of two orthogonal vectors as
\begin{equation} \label{v_result}{\bf
v}=\frac{\talpha(\theta)}{4l}(s_1-s_2)(\nvec_1+\nvec_2)
+\Big[\frac{\beta(\theta)}{2}-\frac{\talpha(\theta)}{4l}(s_1+s_2)\Big](\nvec_2-\nvec_1)\;.
\end{equation}
The physical meaning of Eq.~(\ref{v_result}) can be understood by
considering a specific microscopic model of the motor-filament
coupling, such as the one described in Appendix \ref{app:microscopic}.
In this model the kinematics of two filaments coupled by a motor
cluster is described explicitly in terms of the rate $u(s)$ at which
the cluster steps along the filament and the torsional stiffness
$\kappa$ of the cluster~\cite{torsional_stiffness}.  A comparison of
Eq.~(\ref{v_result}) with Eq.~(\ref{vel_eq2}) assuming a linear
dependence of $u(s)$ on $s$ as $u(s)\sim u_0-u's$, with $u'=-du/ds$,
shows that in the microscopic model $\talpha$ and $\beta$ are
independent of the angle $\theta$, with $\beta=u_0$ and
$\talpha=2lu'$. In general we can identify $\beta$ with the mean rate
at which a motor cluster steps along a filament, i.e., $\beta\sim
(1/l)\int_s u(s)$, while $\talpha$ is controlled by spatial variation
in the stepping rate, with $\talpha\sim 2l \max|du/ds|$. It is then
apparent that the first term on the right hand side of Eq.
(\ref{vrel}) arises from variation in motor activity along the
filament, such as the stalling of motors before detaching upon
reaching a particular point on the filament. It is proportional to the
separation $\bm\xi$ of the filaments' centers of mass and vanishes
when these coincide. The angular dependence of $\talpha$ is chosen so
that this contribution to the relative velocity is largest when
filaments are parallel. The second term in Eq.~(\ref{vrel}),
proportional to $\beta$, vanishes for aligned filaments and drives the
separation or sorting of anti-aligned pairs.

For small angles, we can write the functions $\talpha$ and $\beta$
and $\gamma$ in the form of expansions in powers of
$\nvec_1\cdot\nvec_2$ as
\begin{eqnarray}
\label{alpha}&&
\talpha(\theta)\simeq\talpha_0+\talpha_1(\nvec_1\cdot\nvec_2)\;,\\
\label{beta}&&
\beta(\theta)\simeq\beta_0+\beta_1(\nvec_1\cdot\nvec_2)\;,
\end{eqnarray}
where all coefficients are defined positive. It can be shown that
within the approximation used below, where we only consider  the
first three moments of the filament concentration, no new terms
are obtained in the continuum equations for such moments when
terms of higher order in $\nvec_1\cdot\nvec_2$ are included in
Eqs.~(\ref{alpha}) and (\ref{beta}). Contributions of higher order
in the angle between the filaments only affect the numerical
coefficients of the various parameters in the continuum equations.

\begin{figure}
\center \resizebox{0.49\textwidth}{!}{
  \includegraphics{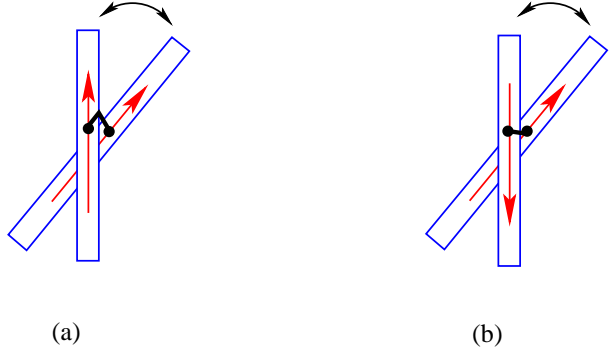}}

\caption{Polar and non-polar clusters interacting with polar
filaments. Assuming that clusters always bind to the smallest
angle, polar clusters ($\gamma_P/\gamma_{NP} \gg 1$) bind only to
filaments in configuration (a) while non-polar clusters
($\gamma_P/\gamma_{NP} \ll 1$) bind to both configurations
equally.}
\label{polar_cluster}       
\end{figure}

The rotational parameter $\gamma$ can be estimated by describing
the crosslinker as a torsional spring of constant $\kappa$, as
shown in Appendix \ref{app:microscopic}, where we find that the
rotational rate induced by a single crosslinker does not depend on
$\theta$. We estimate $\gamma\sim D_r \kappa /k_B T_a$. In this
model we assume that the motor cluster always binds on the side of
the smaller angle between the filaments, as shown in Fig.
\ref{polar_cluster}. We distinguish between polar clusters that
bind preferentially to filaments of the same polarity (Fig.
\ref{polar_cluster}(a)) and nonpolar clusters that bind to
filaments regardless of their relative polarity (Fig.
\ref{polar_cluster}(b)). The probability for such two classes of
protein clusters to bind to filaments will in general depend on
the angle $\theta$ between the filaments, yielding an angular
dependence of the effective rate $\gamma(\theta)$. Again, to lowest order in
$\nvec_1\cdot\nvec_2$ we write
\begin{eqnarray}
\label{gamma}
\gamma(\theta)\simeq\gamma_P+\gamma_{NP}(\nvec_1\cdot\nvec_2)\;,
\end{eqnarray}
The term proportional to $\gamma_P$ favors rotations that align
filaments of the same polarity and describes polar clusters
~\cite{nedelec97,surrey01}, which are in general expected to be active
crosslinks in the presence of ATP. The term proportional to
$\gamma_{NP}$ favors rotation in the direction of angles $\theta<\pi$,
regardless of the relative polarity of the two filaments.  It
describes non-polar clusters which bind to filament pairs of any
orientation~\cite{humphrey02}. Passive cross-linkers (such as
$\alpha$-actinin on F-actin which play a crucial role in the rheology
of actin gels) ~\cite{sackmann96} can be either polar or nonpolar.
Polar clusters ($\gamma_P\not=0$) where not considered in earlier work
by two of us~\cite{TBLMCM03}, but are crucial for the formation of a
polarized phase (see also Ref.~\cite{Aranson05}). Both $\gamma_P$ and
$\gamma_{NP}$ will increase with increasing binding rate of the
clusters to the filament.  It is interesting to speculate that the
kinesin constructs in the experiments by Nedelec et
al.~\cite{nedelec97,surrey01} are polar clusters, while the disordered
myosin II clusters studied by Humphreys et al.~\cite{humphrey02} may
be apolar in nature. We can also imagine that if the binding/unbinding
of the motor clusters does not require ATP, these terms, unlike the
active contributions to the translational currents, would be
independent of the ATP hydrolysis rate.

To determine the net linear velocity ${\bf V}$ and rotational
velocity $\bm{\Omega}=(\bm\omega_1+\bm\omega_2)/2$ we note that
the third law and momentum conservation require that the net
force and torque due to a motor cluster on a pair of filaments
vanishes in the absence of external forces. This yields
\begin{eqnarray}
\label{zeroforce}
&&\zeta_{ij}(\nvec_1)v_{1j}+\zeta_{ij}(\nvec_2)v_{2j}=0\;,\\
\label{zerotorque} &&\zeta_r\bm\omega_1+\zeta_r\bm\omega_2=0\;,
\end{eqnarray}
where
$\zeta_{ij}(\nvec)=\zeta_\parallel\nhat_{i}\nhat_j+\zeta_\perp(\delta_{ij}-\nhat_{i}\nhat_j)$,
with $\zeta_\parallel=k_BT_a/D_\parallel$ and
$\zeta_\perp=k_BT_a/D_\perp$, is the friction tensor of a long
thin rod and $\zeta_r=k_BT_a/D_r$ is the rotational friction. The
vanishing of the net torque on the pair clearly requires
$\bm\omega_2=-\bm\omega_1$, i.e., there is no net rotational
velocity. The net velocity ${\bf V}$ of the pair is generally
nonzero and is given by the solution of Eq.~(\ref{zeroforce}), or
\begin{equation}
\label{netV}
\big[\zeta_{ij}(\nvec_1)+\zeta_{ij}(\nvec_2)\big]V_j=-\frac{1}{2}\big[\zeta_{ij}(\nvec_1)-\zeta_{ij}(\nvec_2)\big]v_j\;.
\end{equation}
Its explicit form is given by
\begin{equation} \label{Vnet_result} {\bf
V}=A(\nvec_1+\nvec_2)+B(\nvec_2-\nvec_1)\;, \end{equation}
with
\begin{eqnarray}
&&A=-\frac{\sigma}{4}~\frac{1-\nvec_1\cdot\nvec_2}
{1-\sigma\nvec_1\cdot\nvec_2}
\Big[\beta(\theta)-\frac{\talpha(\theta)}{2l}(s_1+s_2)\Big]\;,\\
&&B=-\frac{\sigma}{4}~\frac{1+\nvec_1\cdot\nvec_2}
{1+\sigma\nvec_1\cdot\nvec_2}~
\frac{\talpha(\theta)}{2l}(s_1-s_2)\;,
\end{eqnarray}
where
$\sigma=(\zeta_\perp-\zeta_\parallel)/(\zeta_\perp+\zeta_\parallel)>0$.
For long thin rods $\zeta_\perp=2\zeta_\parallel$ and
$\sigma=1/3$.

Equations  (\ref{v_result}) and (\ref{Vnet_result}) display
explicitly the even and odd symmetry of ${\bf V}$ and ${\bf v}$,
respectively, under filament exchange. Note that the net velocity
${\bf V}$ vanishes for isotropic bodies, i.e., when
$\zeta_\parallel=\zeta_\perp$ ($\sigma=0$).

\section{Continuum equations}
\label{sec:Hydrodynamic equations}

Our goal here is to obtain a set of coarse-grained equations to
describe the macroscopic dynamics of active filament mixtures on
scales large compared to the filaments' length, $l$ 
and on timescales long compared to the typical binding times of 
the cross-linkers. 

This level of description is valid when the macroscopic quantities
describing the solution exhibit spatial variations on length scale
much greater than the length of the filaments~\cite{Shimada88}. The
macroscopic quantities we choose to study are the local filament
density, $\rho({\bf r},t)$, the local filament polarization, ${\bf
  p}({\bf r},t)$, and the alignment tensor, $S_{ij}({\bf r},t)$, a
second rank symmetric tensor which measures the local orientational
order in a nematic state. These fields are associated with either
conservation laws (the density) or possible broken continuous
symmetries (${\bf p}$, $S_{ij}$) and therefore control the
hydrodynamic modes of the system. They can be defined as the first
three moments of the distribution $c({\bf r},\nvec,t)$
\cite{TBLMCM03},
\begin{eqnarray}
\label{moments} & & \rho({\bf r},t)=\int_{\nvec}c({\bf\
r},\nvec,t)\;,\nonumber\\
& & {\bf T}({\bf r},t)\equiv\rho(\bfr,t)~{\bf p}({\bf
r},t)=\int_{\nvec}\nvec~c({\bf r},\nvec,t)\;,\nonumber\\
& & Q_{ij}({\bf r},t)\equiv\rho(\bfr,t)~S_{ij}({\bf
r},t)=\int_{\nvec} \hat{Q}_{ij}(\nvec)~c({\bf
r},\nvec,t)\;,
\end{eqnarray}
with $\hat{Q}_{ij}(\nvec)=\nhat_i\nhat_j-\frac{1}{2}\delta_{ij}$.
Hydrodynamic equations for these coarse-grained densities can  be
obtained by writing an exact moment expansion for $c({\bf
r},\nvec,t)$ (see Appendix \ref{app:expansion}) and
truncating this expansion at the third moment. To derive the
continuum equations we assume that all quantities are slowly
varying on the scales of interest and expand the concentration of
filaments $c({\bf r}_1+\bm\xi,\nvec_2)$ and the
crosslinker density $m({\bf r}_1+\nvec_1s_1)$ in the expressions
for the active currents near their values at ${\bf r}_1$ as
\begin{eqnarray}
\label{c_exp} c({\bf r}_1+\bm\xi,\nvec_2)&= &c({\bf
r}_1,\nvec_2)+\xi_i\partial_{1i}c({\bf
r}_1,\nvec_2)+\frac{1}{2}\xi_i\xi_j\partial_{1i}\partial_{1j}c({\bf
r}_1,\nvec_2)+{\cal O}(\nabla^3)\;,
\end{eqnarray}
\begin{eqnarray}
\label{m_exp} m(\bfr_1+\nvec_1s_1)&=
&m(\bfr_1)+\nhat_{1i}s_1\partial_{1i}m(\bfr_1)
+\frac{1}{2}\nhat_{1i}\nhat_{1j}s^2_1\partial_{1i}\partial_{1j}m(\bf
r_1)+{\cal O}(\nabla^3)\;.
\end{eqnarray}
When the expansions (\ref{c_exp}) and (\ref{m_exp}) are inserted
in Eqs.~(\ref{Jtact}) and (\ref{Jract}), the integration over
$s_1$ and $s_2$ can be carried out term by term. An analogous
expansion is used to approximately evaluate the excluded volume
interaction, as well as in the equation for the motor
concentration. Some details of the derivation of the hydrodynamic
equations for the motor density, filament density, polarization
and alignment tensor are given in Appendix \ref{app:currents}.

For simplicity, here we consider the nonlinear continuum equations
retaining only terms up to second order in the gradients.  While
the analysis of the linear stability of homogeneous states with
terms up to fourth order in the gradients does introduce a new
length scale (see Appendix \ref{app:fourth_order}), the simplified
equations  are instructive and capable of describing much of the
physics. The motor density obeys a simple diffusion equation given
by
\begin{eqnarray}
\label{motor_eqn} \partial_tm_a
&=&D_m\nabla^2m_a-\bnabla\cdot(m_a\bfT)\;,
\end{eqnarray}
where the second term describes convection of the motors along the
filaments~\cite{Lee01}. The equations for the filament density,
polarization and alignment tensor are
\begin{equation}
\label{rhoeqn_impl}
\partial_t\rho=-\partial_i J_i\;,
\end{equation}
\begin{eqnarray}
\label{peqn} \partial_t(\rho p_i)&=&-\partial_j J_{ij}-R_i\;,
\end{eqnarray}
\begin{eqnarray}
\label{Qeqn} \partial_t(\rho S_{ij})&=&-\partial_k
J_{ijk}-R_{ij}\;,
\end{eqnarray}
where the currents are given by
\begin{eqnarray}
\label{Jr}J_i({\bf r},t)&=&\int_\nvec  J_{ci}(\nvec,{\bf r},t)\;,
\end{eqnarray}
\begin{eqnarray}
\label{JT}J_{ij}({\bf r},t)&=&\int_\nvec  \nhat_i J_{cj}(\nvec,{\bf
r},t)\;,
\end{eqnarray}
\begin{eqnarray}
\label{JQ}J_{ijk}({\bf r},t)=\int_\nvec
\hat{Q}_{ij}J_{ck}(\nvec,{\bf r},t)\;.
\end{eqnarray}
The rotational current does not contribute to the density
equation, but it yields the source terms $R_i$ and $R_{ij}$ in the
equations for the polarization and alignment tensor. These are given by
\begin{eqnarray}
\label{RT}R_i({\bf r},t)=\int_\nvec \nhat_i~ \bm{\mathcal R} \cdot
\bm{\mathcal J}_c(\nvec,{\bf r},t)\;,
\end{eqnarray}
\begin{eqnarray}
\label{RQ}R_{ij}({\bf r},t)=\int_\nvec \hat{Q}_{ij}~ \bm{\mathcal
R} \cdot \bm{\mathcal J}_c(\nvec,{\bf r},t)\;.
\end{eqnarray}
The explicit form of the translational (\ref{Jr}-\ref{JQ}) and
rotational (\ref{RT}-\ref{RQ}) currents is given in
Appendix \ref{app:currents}. The equation for the density $\rho$
has the form of a continuity equation, as required by filament
number conservation. The local polarization ${\bf p}$ and the
alignment tensor $S_{ij}$ define the order parameters needed to
characterize the ordered states of the system and are not
conserved variables. Each ordered state discussed below will,
however, be characterized by a broken orientational symmetry and a
corresponding broken symmetry variable (a unit vector along the
direction of broken symmetry) whose fluctuations are infinitely
long lived at large wavelength, as required for hydrodynamic
modes.

\section{Homogeneous States}
\label{sec:Homogeneous_states}

We begin by identifying the possible homogeneous steady states of
the system. In this case all contributions to the dynamical
equations for the filament solution come from the rotational
currents. Both the motor and filament densities have constant
values $\tm=mb^2$ and  $\rho=\rho_0$. The equation for the
polarization and the alignment tensor are given by
\begin{eqnarray}
\label{st_P_dim}&&
\partial_tp_i=-\big[D_r-\gamma_P\tm\rho_0\big]p_i+
\Big[4D_r\rho_0/\rho_N+(\gamma_{NP}-2\gamma_P)\tm\rho_0\Big]
S_{ij}p_j\;,
\end{eqnarray}
\begin{eqnarray}
\label{st_Q_dim}&&\partial_tS_{ij}=-\Big[4D_r(1-\rho_0/\rho_N)-\gamma_{NP}\tm\rho_0\Big]S_{ij}
+2\gamma_P\tm\rho_0 \Big(p_ip_j-\frac{1}{2}p^2\Big)\;,
\end{eqnarray}
where all filament densities are measured in units of $l^2$, and
$\rho_N=3\pi/2$. The motor-induced rotational rates $\gamma_P$ and
$\gamma_{NP}$ have dimensions of frequency and represent the effect of
polar and nonpolar motor clusters, respectively. For simplicity we
denote by $\tm$ the total dimensionless density of crosslinkers,
without distinguishing between stationary and active protein clusters.
One can imagine situations, however, where $\gamma_P$ will in general
be proportional to the ATP consumption rate, but the nonpolar coupling
$\gamma_{NP}$ will be only weakly affected by ATP concentration. In
the following all lengths are measured in units of the filament length
$l$ and times are measured in units of $D_r^{-1}$.

There are three possible homogeneous stationary states for the
system, obtained by solving Eqs.~(\ref{st_P_dim}) and
(\ref{st_Q_dim}) with $\partial_tp_i=0$ and $\partial_t S_{ij}=0$.
These are:
\begin{eqnarray}
&&{\rm isotropic}\hspace{0.1in}{\rm state}\hspace{0.1in}{\rm
(I):}\hspace{0.3in}
p_i=0\hspace{0.2in}S_{ij}=0\;,\nonumber\\
&&{\rm nematic}\hspace{0.1in}{\rm state}\hspace{0.1in}{\rm
(N):}\hspace{0.3in}
p_i=0\hspace{0.2in}S_{ij}\not=0\;,\nonumber\\
&&{\rm polarized}\hspace{0.1in}{\rm state}\hspace{0.1in}{\rm
(P):}\hspace{0.3in}
p_i\not=0\hspace{0.2in}S_{ij}\not=0\;.\nonumber
\end{eqnarray}
At low density the only solution is $p_i=0$ and $S_{ij}=0$ and the
system is isotropic (I). The homogeneous isotropic state can
become unstable at high filament and/or motor density, as
described below.

To discuss the instabilities it is convenient to rewrite
Eqs.~(\ref{st_P_dim}) and (\ref{st_Q_dim}) in a more compact form
as
\begin{eqnarray}
\label{st_P}&& \partial_tp_i=-a_1p_i+b_1\rho_0 S_{ij}p_j\;,
\end{eqnarray}
\begin{eqnarray}
\label{st_Q}&&\partial_tS_{ij}=-a_2S_{ij}+b_2\rho_0
\Big(p_ip_j-\frac{1}{2}p^2\Big)\;.
\end{eqnarray}
The coefficients $a_1$, $b_1$, $a_2$, and $b_2$ are given by
\begin{eqnarray}
&& a_1=1-\tm\gamma_P\rho_0/D_r\;,\\
&&a_2=4\big[1-\rho_0/\rho_N-\gamma_{NP}\tm\rho_0/(4D_r)\big]\;,\\
&&b_1=4\big[\rho_N^{-1}+(\gamma_{NP}-2\gamma_P)\tm/(4D_r)\big]\;,\\
&&b_2=2\gamma_P\tm/D_r\;.
\end{eqnarray}

In the absence of crosslinkers ($\gamma_P=\gamma_{NP}=0$),
no homogeneous polarized state with a nonzero mean value of ${\bf
p}$ is obtained. There is, however, a transition at the density
$\rho_N= 3\pi/2$ from an isotropic state with $S_{ij}=0$ to a
nematic state with $S_{ij}=S_0(n_in_j-\frac{1}{2}\delta_{ij})$,
with ${\bf n}$ a unit vector along the direction of broken
symmetry. The transition here is identified with the
change in sign of the coefficient $a_2$ of $S_{ij}$ on the right
hand side of Eq.~(\ref{st_Q}). A negative value of $a_2$ that
controls the decay rate of $S_{ij}$ signals an instability of the
isotropic homogeneous state. This occurs when excluded volume
effects dominate at $\rho_0=\rho_N$. The homogeneous state is
isotropic for $\rho_0<\rho_N$ and nematic for $\rho_0>\rho_N$.
A mean-field description of such a transition, which is
continuous in 2d (but first order in 3d), requires that one
incorporates cubic terms in $S_{ij}$ in the
equation for the alignment tensor. Adding a term $-c_2 \rho_0^2
S_{kl}S_{kl}S_{ij}$ to Eq.~(\ref{st_Q}) we obtain $S_0 =
\frac{1}{\rho_0}\sqrt{-2a_2/c_2}
=\frac{1}{\rho_0}\sqrt{-8(1-\rho_0/\rho_{N})/c_2}$.

\begin{figure}
\center \resizebox{0.49\textwidth}{!}{
  \includegraphics{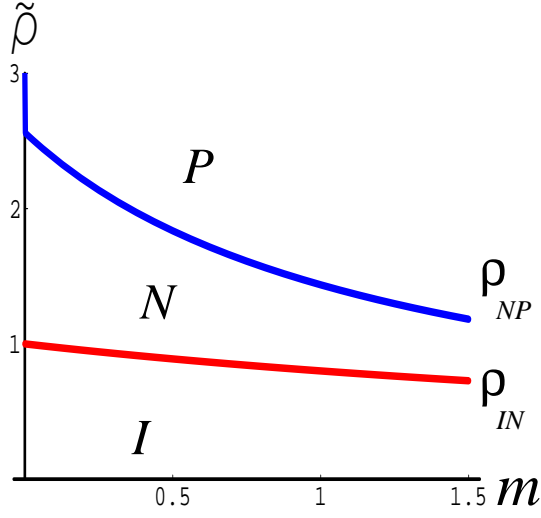}
}

\caption{
The homogeneous phase diagram for $g<1/4$. For all values
of $\tm$ a region of nematic phase exists between the isotropic
and polarized phases ($\gamma_P/D_r=1$, $g=1/10$ and $c_2=50$).}
\label{PD1}       
\end{figure}

If $\gamma_P=0$, but $\gamma_{NP}\not=0$,  there is again no
stable polarized state. The presence of a concentration of
nonpolar crosslinkers does, however, renormalize the
isotropic-nematic (IN) transition, which occurs at a lower
filament density given by
\begin{equation}
\label{rhoIN}
\rho_{IN}(\tm)=\frac{\rho_N}{1+\tm\gamma_{NP}\rho_N/(4D_r)};.
\end{equation}
The presence on nonpolar crosslinks favors filament alignment and
shifts $\rho_{IN}$ downward, as shown in Fig.~\ref{PD1}. A
qualitatively similar result has been obtained in numerical
simulation of a two-dimensional system of rigid filaments
interacting with motor proteins grafted to a substrate~\cite{kraikivski}. In this
case the motors promote alignment by exerting longitudinal forces
on the filaments. The amount of nematic order $S_0$ is also
enhanced by motor activity, with  $S_0
=\frac{1}{\rho_0}\sqrt{-2a_2/c_2}=
\frac{1}{\rho_0}\sqrt{-8(1-\rho_0/\rho_{IN}(\tm))/c_2}$.

If $\gamma_P$ is finite, the system can order in both polarized and
nematic homogeneous states. The homogeneous isotropic state can
become unstable in two ways. As in the  case $\gamma_P=0$, a
change in sign of the coefficient $a_2$ signals the transition to
a nematic (N) state at the density $\rho_{IN}(\tm)$ given in
Eq.~(\ref{rhoIN}). In addition, the isotropic state can become
linearly unstable via the growth of polarization fluctuations in
any arbitrary direction. This occurs above a second critical
filament density,
\begin{equation}
\label{rhoIP}\rho_{IP}(\tm)=\frac{D_r}{\gamma_P\tm}\;,
\end{equation}
 defined
by the change in sign of the coefficient $a_1$ controlling the
decay of polarization fluctuations in Eq.~(\ref{st_P}). For $\rho_0>\rho_{IP}(\tm)$
the homogeneous state is polarized (P), with ${\bf p}\not=0$. The
alignment tensor also has a nonzero mean value in the polarized
state as it is slaved to the polarization. The location of the
boundaries between the various homogeneous states is controlled by
the ratio $g=\gamma_P/\gamma_{NP}$ that measures the polarity of
motor clusters. One can identify two scenarios depending on the
value of $g$.

I) For $g < 1/4$, the density $\rho_{IP}$ is always larger than
$\rho_{IN}$   and a region of nematic phase exists for all values
of $\tm$. At sufficiently high filament and motor densities, the
nematic state becomes unstable. To see this, we linearize
Eqs.~(\ref{st_P}) and (\ref{st_Q}) by letting
$S_{ij}=S_{ij}^0+\delta S_{ij}$ and $\delta p_i=p_i$. Fluctuations
in the alignment tensor are uniformly stable for $a_2<0$, but
polarization fluctuations along the direction of broken symmetry
become unstable for $a_1\leq\rho_0 b_1 S_0/2$, i.e., above a
critical density
\begin{equation}
\rho_{NP}=\frac{D_r}{\tm g\gamma_P} \Big[ 1 + {b_1^2 \over c_2
R}\Big( 1 - \sqrt{1 + {2 c_2 R (1-R)\over b_1^2}}\Big)\Big]
\end{equation}
where $R = \rho_{IN} / \rho_{IP}$. The polarized state at
$\rho_0>\rho_{NP}$ has
\begin{eqnarray}
&&p_i^0=p_0\hat{p}_i\;,\\
&&S_{ij}^0=S_P(\hat{p}_i\hat{p}_j-\delta_{ij}/2)\;,
\end{eqnarray}
with ${\bf \hat{p}}$ a unit vector in the direction of broken
symmetry and
\begin{eqnarray}
&&p_0^2=\frac{2a_1a_2}{\rho_0^2
  b_1b_2}\Big[1-\Big(\frac{2a_1}{b_1\rho_0S_0}\Big)^2\Big]\;,\\
&&S_P=S_0\sqrt{1-\frac{\rho_0^2
b_1b_2}{2a_1a_2}p_0^2}=2\Big|\frac{a_1}{\rho_0b_1}\Big|\;.
\end{eqnarray}
The  "phase diagram" for $g<1/4$ is shown in
Fig.~\ref{PD1}.

\begin{figure}
\center \resizebox{0.49\textwidth}{!}{
  \includegraphics{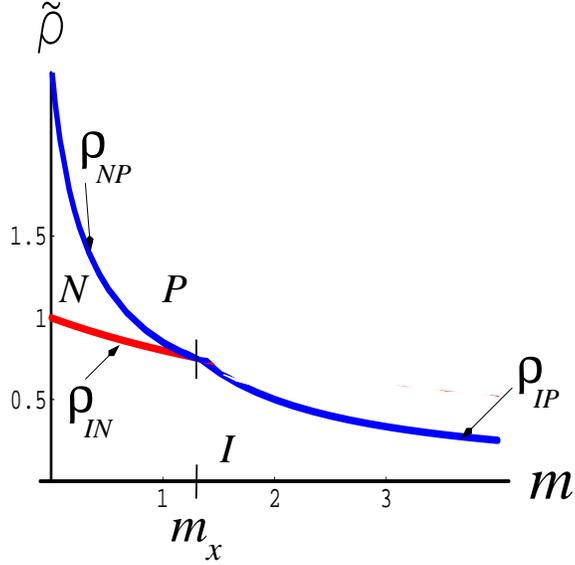}
}

\caption{
The phase diagram for $g>1/4$. For $\tm>\tm_x$, where
$\rho_{IN}$ and $\rho_{IP}$ intersect, no N state exists and the
system goes directly from the I to the P state ($\gamma_P/D_r=1$,
$g=1$ and $c_2=50$).}
\label{PD2}       
\end{figure}

II) When $g > 1/4$, the boundaries for the I-N and the N-P
transitions cross at
\begin{equation}
\label{tmx} \tm_x=\frac{\rho_N D_r/\gamma_P}{1-1/(4g)}\;,
\end{equation}
where $\rho_{IN}=\rho_{IP}=\rho_{NP}$ and the phase diagram has
the topology shown in Fig.~\ref{PD2}. For $\tm>\tm_x$ the polarity
of motor clusters renders the nematic state unstable at all
densities larger than $\rho_{IN}(\tm)$ and the system goes
directly from the I to the P state at $\rho_{IP}$, without an
intervening N state. At the onset of the polarized state the
alignment tensor is again slaved to the polarization field,
\begin{math}
S_{ij}=\frac{b_2}{a_2}\rho_0~(p_ip_j-\frac{1}{2}\delta_{ij}p^2)\;,
\end{math}
and ${\bf p}=p_0{\bf \hat{p}}$. The value of $p_0$ is determined
by cubic terms in Eq. (\ref{st_P}) not included here.

Finally, we note that if $\gamma_{NP}=0$, with $\gamma_P\not=0$
(i.e., $g\rightarrow\infty$), the I-N transition is independent of
motor density and always occurs at $\rho_0=\rho_N$. The motor
density where $\rho_{IN}=\rho_{IP}$ reduces to
$\tm_x=\rho_ND_r/\gamma_P$. 


Estimates of the various parameters can be obtained using a
microscopic model of the motor-filament interaction of the type
described in Appendix~\ref{app:microscopic}. Using parameter values
appropriate for kinesin ($\kappa\sim 10^{-22}\mbox{nm/rad}$
\cite{howard}) we estimate
$\gamma_P\sim\gamma_{NP}\sim\kappa/\zeta_r=\kappa D_r/(k_BT_a)\sim
10^{-1}\mbox{sec}^{-1}$, where we used the value $D_r\sim
10^{-2}\mbox{sec}^{-1}$ appropriate for long thin rods in an aqueous
solution~\cite{Hunt93} and $T_a\sim 300\, \mbox{K}$. 
Using $l\sim 10\mu \mbox{m}$, $b\sim 10\, \mbox{nm}$,
the value $\tm_x$ above which no nematic state exist is found to
correspond to a three-dimensional crosslinker density of about
$0.5-1\mu \mbox{M}$ for $g=1$ and a sample thickness of order $1\mu \mbox{m}$.
This value is of order of the motor densities used in experiments
on purified microtubule-kinesin mictures such as those of Ref.
\cite{humphrey02}, suggesting that the filament solution in this
experiments is always in the polarized state.


\section{Dynamics of inhomogeneous states}
\label{sec:inhomogeneous}

Experiments on motor/filament mixtures have shown that uniform
states are often unstable to the formation of complex spatial
structures. These instabilities arise from the growth of spatial
fluctuations in the hydrodynamic
fields~\cite{takiguchi,urrutia,nedelec97,surrey01}.  In particular, the
rate of motor-induced filament bundling can exceed that of
filament diffusion yielding the unstable growth of density
inhomogeneities~\cite{nakazawa96,SNbook,Kruse00,Kruse03,Kruse01,TBLMCM03}. States
with spatially varying orientational order, where the filaments
spontaneously arrange in vortex and aster structures, are also
possible~\cite{Lee01,Sankararaman03,Aranson05}. To understand the
different nature of the instability from each homogeneous state,
we now examine the dynamics of spatially-varying fluctuations in
the hydrodynamic fields in each of the stationary homogeneous
states of the system. The hydrodynamic fields are those with
characteristic decay times that exceed any microscopic relaxation
time and become infinitely long at long wavelengths. We find that
the low frequency hydrodynamic modes of this active system are
determined by fluctuations in the conserved densities and in
variables associated with broken symmetries.  A change in sign in
the decay rate of these modes signals an instability of the
macroscopic state of interest. For simplicity we only discuss here
the case of constant motor density. In addition in this section we
let $\alpha_0=\talpha_0/48$, $\alpha_1=\talpha_1/48$ and
assume $\alpha_0=\alpha_1=\alpha$, $\beta_0=\beta_1=\beta$.
This approximation is justified by
the estimate for these motor-induced parameters obtained in
Appendix \ref{app:microscopic} using a microscopic model for the
the motor-mediated filament interaction~\cite{TBLMCM05}. We
consider separately the spatially varying hydrodynamic modes in
each of the homogeneous states: isotropic, nematic and polarized
phases.

\subsection{Isotropic state}
\label{subsec:isotropic}

The isotropic homogeneous state has $\rho=\rho_0$, ${\bf p}_0=0$ and
$S^0_{ij}=0$. The only hydrodynamic variable in this state is the
density of filaments. The polarization vector ${\bf p}$ and nematic
orientation tensor $S_{ij}$ are {\em not} hydrodynamic modes and
therefore relax to zero on microscopic timescales.  However, 
due to dynamical constraints such as entanglements 
the relaxation of ${\bf p}$ and $S_{ij}$ can become sufficiently 
slow to yield finite lifetime and finite wavelength inhomogeneities~\cite{TBLMCM03,ziebert}.

The nonlinear equation for the density is given by
\begin{equation}
\partial_t \rho = {3 D \over 4} \bnabla\cdot(1+v_0\rho)\bnabla\rho -
\alpha\bnabla\cdot(\rho\bnabla\rho)\;,
\end{equation}
where $v_0=2/\pi$.
The active current proportional to $\alpha$ has an effect opposite
to that of thermal diffusion as it tends to build up density
inhomogeneities. As we will see below this term drives filament
bundling and is the main pattern-forming mechanism in each of the
homogeneous states.

\subsubsection{Linear Stability}

To examine the stability of the isotropic state we consider the
dynamics of fluctuations $\delta\rho(\bfr)=\rho(\bfr)-\rho_0$ of the
density about its steady-state value, $\rho_0$ to {\em linear} order
in $\delta \rho$. By expanding the fluctuations in Fourier space,
\begin{equation}
\label{fourier_rho} \delta\rho(\bfr)=\Sigma_{\bfk}\rho_{\bfk}e^{i\bfk\cdot\bfr}\;,
\end{equation}
the linearized equation for the Fourier amplitudes is given by
\begin{eqnarray}
\partial_t\rho_{\bf k}&=&-k^2\Big[\frac{3}{4}D(1+\rho_0v_0)-\alpha\tm\rho_0\Big]\rho_{\bf
k}\;.
\end{eqnarray}
The relaxation of density fluctuations is governed by a diffusive
mode of frequency
\begin{equation}
z_\rho(k)=-k^2\Big[\frac{3}{4}D(1+\rho_0v_0)-\alpha\tm\rho_0\Big]\;.
\end{equation}
The isotropic state becomes  unstable against the growth of
density fluctuations if $z_\rho(k)>0$, or $\alpha>\alpha_{c}$,
with
\begin{equation}
\alpha_{c}=\frac{3D(1+\rho_0v_0)}{4\tm\rho_0}\;.
\end{equation}
Conversely, the homogeneous isotropic state becomes unstable for
filament densities larger than a critical value
\begin{equation}
\label{rhoB} \rho_B^I=\frac{3D}{4\tm\alpha-3Dv_0}\sim
\frac{3D}{4\tm\alpha}\;. \end{equation}
This bundling instability of the isotropic state has been
discussed elsewhere~\cite{TBLMCM03,TBLMCM05}. A proper description
of the bundling instability requires that one incorporates terms
up to $4^{th}$ order in the gradient expansion of the hydrodynamic
fields~\cite{ziebert}. The $4^{th}$ order terms introduce a new
length scale above which the homogeneous state becomes again
stable, as shown in Appendix \ref{app:fourth_order}. The onset of
the instability is, however, controlled entirely by the quadratic
terms considered here.

\subsection{Nematic state}
\label{subsec:nematic}

The  homogeneous nematic state is characterized by $\rho=\rho_0$,
${\bf p}_0=0$, and  $S_{ij}=S_0(n_in_j-\frac{1}{2}\delta_{ij})$,
where ${\bf n}$ is
 a unit vector in the direction of broken
symmetry, known as the director field. For concreteness we choose
${\bf n}={\bf \hat{y}}$. The hydrodynamic fields of such an
overdamped nematic liquid crystal are the density and the
director. The symmetry of the nematic state requires that the
hydrodynamic equations incorporate the symmetry ${\bf
n}\rightarrow -{\bf n}$. The magnitude $S_0$ of the alignment
tensor is not a hydrodynamic field and will be assumed constant
below. For simplicity we also neglect excluded volume corrections.
The nonlinear hydrodynamic equations for filament density and
director field are then given by
\begin{eqnarray}
\label{rho_nematic}
\partial_t\rho&=&\frac{3D}{4}\Big(1-\frac{S_0}{3}\Big)\nabla^2\rho-\frac{1}{2}(1-S_0)\alpha\tm\nabla^2\rho^2
+\frac{S_0}{2}\partial_i\Big[\big(D-4\alpha\tm\rho\big)n_in_j\partial_j\rho\Big]\nonumber\\
&&
+\frac{S_0}{2}\partial_i\Big[\big(D-\frac{4}{3}\alpha\tm\rho\big)\rho\partial_j(n_in_j)\Big]\;,
\end{eqnarray}
%
%
%

%
\begin{eqnarray}
\label{director_nematic}
\partial_t n_i &=&
\frac{1}{6}\delta_{ij}^T\Big\{\Big[\frac{7D}{2\rho}+\frac{1}{8}\Big(1-\frac{S_0}{2}\Big)\gamma_{NP}\tm\Big]\nabla^2(\rho n_j)
+\Big[\frac{D}{\rho}+\frac{1}{8}(1-S_0)\gamma_{NP}\tm\Big]\partial_j\bm\nabla\cdot(\rho{\bf n})\nonumber\\
&&+\Big[\frac{D}{\rho}+\frac{1}{8}(1+2S_0)\gamma_{NP}\tm\Big]n_kn_l\partial_l\partial_k(\rho n_j)
-\Big[\frac{D}{\rho}\Big(1-\frac{3}{4S_0}\Big)
+\frac{1}{8}\Big(1-\frac{1}{2S_0}-\frac{S_0}{2}\Big)\gamma_{NP}\tm\Big]n_k\partial_k\partial_j\rho\nonumber\\
&&+\Big[\frac{D}{\rho}+\frac{1}{8}\Big(1+\frac{S_0}{3}\Big)\gamma_{NP}\tm\Big]\rho\partial_k(n_kn_l)\partial_l n_j
-\Big[\frac{D}{\rho}+\frac{1}{8}\Big(1-\frac{S_0}{3}\Big)\gamma_{NP}\tm\Big]\rho n_k(\partial_kn_l)\partial_jn_l\Big\}\nonumber\\
&&-\frac{1}{9}\delta_{ij}^T\alpha\tm\Big\{\frac{3}{2S_0\rho}n_k\partial_k\partial_j\rho^2+5S_0\rho\partial_k(n_kn_l)\partial_ln_j
+2S_0\rho n_l(\partial_ln_j)\partial_kn_k\nonumber\\
&&+\Big[(5-3S_0)\partial_kn_j+\frac{7}{4}\partial_jn_k+\frac{9}{4}\delta_{jk}\partial_ln_l
+2(2+3S_0)n_kn_l\partial_ln_j\Big]\partial_k\rho\Big\}\;.
\end{eqnarray}
where $\delta_{ij}^T=\delta_{ij}-n_in_j$ projects in the direction transverse to ${\bf n}$.

In the density equation, Eq.~(\ref{rho_nematic}), activity
plays the same role as in the isotropic state, with
bundling ($\sim\alpha$) opposing diffusion and
eventually driving the instability of the homogeneous state, as
described below.  The first and second terms on the right hand
side of Eq.~(\ref{director_nematic}) for the nematic director are
the elastic restoring forces associated with bend and splay
deformations, respectively. These elastic constants are
softened by filament bundling, while motor-induced alignment
($\sim\gamma_{NP}$) tends to stabilize the homogeneous nematic state.
A solution with $\rho={\rm constant}$ requires
vanishing of splay, i.e., $\bm\nabla\cdot{\bf n}=0$. In this case the
director equation reduces to
\begin{eqnarray}
\label{director_nematic_constant_rho}
\partial_t n_i &=&
\frac{1}{6}\delta_{ij}^T\Big\{\Big[\frac{7D}{2}+\frac{1}{8}\Big(1-\frac{S_0}{2}\Big)\gamma_{NP}\tm\rho\Big]\nabla^2 n_j
+\Big[D+\frac{1}{8}(1+2S_0)\gamma_{NP}\tm\rho\Big]n_kn_l\partial_l\partial_k n_j\nonumber\\
&&+\frac{S_0}{24}\gamma_{NP}\tm\rho n_k(\partial_k n_l)[\partial_j n_l+\partial_l n_j]
-\frac{10S_0}{3}\alpha\tm\rho n_k(\partial_k n_l)(\partial_l n_j)\Big\}\;.
\end{eqnarray}
In this case bundling does not play any role to linear order.

To discuss the stability of the homogeneous nematic state, we
consider the dynamics of spatially varying fluctuations of the
hydrodynamic fields about their mean values, by letting
\begin{eqnarray}
&&\rho(\bfr)=\rho_0+\delta\rho(\bfr)\;,\\
&&{\bf n}(\bfr)={\bf \hat{y}}+\delta{\bf n}_\perp(\bfr)\;.
\end{eqnarray}
To lowest order in the fluctuations  $\delta{\bf n}_\perp$ is
perpendicular to ${\bf \hat{y}}$, i.e., in the two-dimensional
geometry considered here, $\delta{\bf n}_\perp=\delta n_x\bf
\hat{x}$. The linearized equation for the Fourier amplitude of
density and director fluctuations for $S_0=1$ are given by
\begin{eqnarray}
\partial_t \rho_{\bf k}&=&-\frac{1}{2}\Big[Dk^2+(D-4\alpha\tm\rho_0)k_y^2\Big]\rho_\bfk
-\Big(D-\frac{4}{3}\alpha\tm\rho_0\Big)\rho_0 k_xk_y n_{\bf k}\;,
\end{eqnarray}
\begin{eqnarray}
\partial_t n_{\bf k}&=&-\frac{1}{4}\Big[\Big(3D+\frac{1}{24}\gamma_{NP}\tm\rho_0\Big)k^2+\frac{1}{4}\gamma_{NP}\tm\rho_0 k_y^2\Big]n_{\bf k}
-\frac{1}{8}\Big(D-\frac{8}{3}\alpha\tm\rho_0\Big)k_xk_y\frac{\rho_\bfk}{\rho_0}\;.
\end{eqnarray}

The hydrodynamic modes in the nematic state describe the coupled
decay of density and director fluctuations. They are always
diffusive and are given by
\begin{eqnarray}
\label{Nmodes} z_\pm(k,\phi)&=&-{\cal D}^N_{\pm}(\phi)k^2\;,
\end{eqnarray}
where
\begin{eqnarray}
{\cal D}^N_{\pm}(\phi)&=&
\frac{1}{4}\Big\{\frac{5}{2}D+\frac{1}{48}\gamma_{NP}\tm\rho_0
+\Big(D+\frac{1}{8}\gamma_{NP}\tm\rho_0-4\alpha\tm\rho_0\Big)\cos^2\phi\Big\}\nonumber\\
&&\mp
\frac{1}{4}\Big\{\Big[\frac{1}{2}D+\frac{1}{48}\gamma_{NP}\tm\rho_0
-\Big(D-\frac{1}{8}\gamma_{NP}\tm\rho_0-4\alpha\tm\rho_0\Big)\cos^2\phi\Big]^2\nonumber\\
&&+2\Big(D-\frac{4}{3}\alpha\tm\rho_0\Big)\Big(D-\frac{8}{3}\alpha\tm\rho_0\Big)\sin^2\phi\cos^2\phi\Big\}^{1/2}\;,
\end{eqnarray}
and $\phi$ is the angle between the wavevector ${\bf k}$ and the
direction of the broken symmetry ($\bf \hat{y}$).
To gain some insight in the angular dependence of the modes it is
useful to consider the behavior for special directions of the
wavector. For wavevectors ${\bf k}$ along the $\bf \hat{y}$
direction ($\phi=0$), density and director fluctuations decouple
and we obtain
\begin{eqnarray}
&&z_\rho(k_y)=-\Big[D-2\alpha\tm\rho_0\Big]k_y^2\;,\\
&&z_n(k_y)=-\frac{1}{4}\Big[3D+\frac{7}{24}\gamma_{NP}\tm\rho_0\Big]k_y^2\;.
\end{eqnarray}
In this case the director fluctuations are always stable, while
the density fluctuations become unstable for filament densities
above a critical value $\rho_B^N$, given by
\begin{eqnarray}
\label{rhoBNzero} \rho_B^N(\phi=0)= \frac{D}{2\tm\alpha}\;.
\end{eqnarray}
For ${\bf k}$ along $\bf \hat{x}$ ($\phi=\pi/2$) density and
director fluctuations again decouple, but both eigenvalues are
always negative, with
\begin{eqnarray}
&&z_\rho(k_x)=-\frac{1}{2}Dk_x^2\;,\\
&&z_n(k_x)=-\frac{1}{4}\Big[3D+\frac{1}{24}\gamma_{NP}\tm\rho_0\Big]k_x^2\;.
\end{eqnarray}
The homogeneous nematic state is linearly stable for all parameter
values  against long-wavelength fluctuations that only exhibit
spatial variation in the direction normal to that of the mean
filament orientation.

\begin{figure}
\center \resizebox{0.7\textwidth}{!}{
  \includegraphics{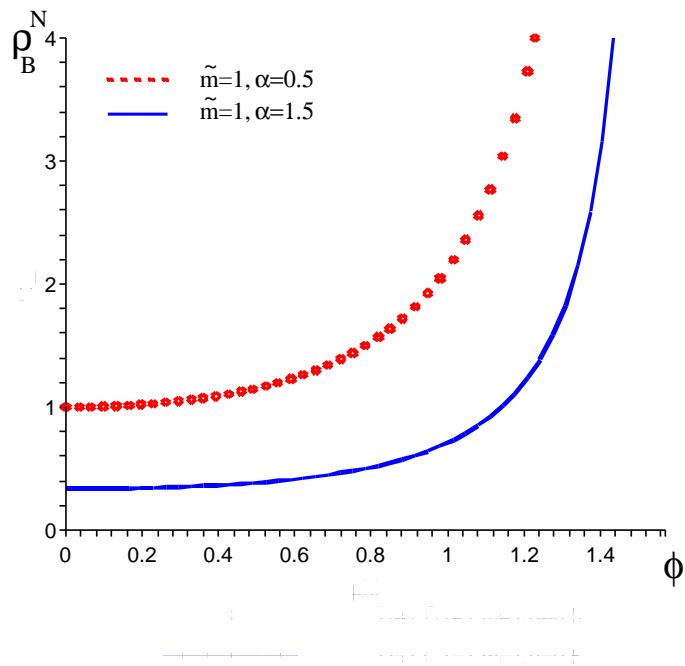}
}

\caption{(color online) The critical filament density $\rho_{B}^N(\phi)$ where
the homogeneous nematic state becomes linearly unstable is shown
as a function of the angle $\phi$ between the wavevector ${\bf k}$
and the direction of broken symmetry for $D=1,~\gamma_{NP}=1,~\tm=1$ and
two values of $\alpha$. At $\phi=0$
the critical density is given by Eq.~(\ref{rhoBNzero}). When
either $\alpha$ or $\tm$ are increased, the density
$\rho_B^N(\phi)$ shifts to lower values at all angles
and the region of stability of the homogeneous nematic state
shrinks.}
\label{rho_N}       
\end{figure}

In general the critical filament density $\rho_{B}^N(\phi)$ above which the
homogeneous nematic state is unstable has a
complicated angular dependence. It increases with $\phi$ and it
diverges for $\phi\rightarrow\pi/2$, where the homogeneous state
is linearly stable for all filament density. Its angular dependence is
shown in Fig.~\ref{rho_N} for a few values of parameters.

For all angles the instability is controlled by the bundling rate,
$\alpha$, while the rotational rate $\gamma_{NP}$ always tends to
stabilize the homogeneous nematic state.

\subsection{Polarized homogeneous state}
\label{subsec:polarized}

The homogeneous polarized state is characterized by $\rho=\rho_0$,
${\bf p}_0=p_0 \pvec$, and $S_{ij}^0=S_P(\phat_i
\phat_j-\frac{1}{2}\delta_{ij})$, with $\pvec$ a unit vector
pointing along the direction of broken symmetry. The hydrodynamic
fields are the filament density and the unit vector $\pvec$. The
magnitude $p_0$ of the polarization is not a hydrodynamic field
and will be assumed constant in the following. In the polarized
state the alignment tensor is slaved to the polarization and does
not yield any additional hydrodynamic field. Assuming for
simplicity $p_0=1$ and neglecting excluded volume corrections, the nonlinear
hydrodynamic equations for filament density and polarization
direction are given by~\cite{nonlinear_polarized}
\begin{eqnarray}
\label{rhoeqn_pol_state}
\partial_t\rho-\frac{7}{36}\beta\tm\bm\nabla\cdot(\rho^2\pvec)
=\frac{3}{4}D\nabla^2\rho-\frac{3}{4}\alpha\tm\nabla^2\rho^2
-\frac{1}{2}\alpha\tm\partial_i\partial_j(\rho^2\phat_i\phat_j)\;,
\end{eqnarray}
and
\begin{eqnarray}
\label{phateqn_pol_state}
\Big[\partial_t+\frac{\tm}{36}\beta\rho\pvec\cdot\bm\nabla\Big]\phat_i&=&
\frac{13\tm}{36}\beta\delta_{ij}^T\partial_j\rho
+\frac{1}{96}\gamma_{NP}\tm\delta_{ij}^T\phat_k\partial_k\partial_j\rho\nonumber\\
&&+\delta_{ij}^T\Big[\Big(\frac{5D}{8\rho}+\frac{\gamma_P\tm}{24}\Big)\nabla^2(\rho\phat_j)
+\frac{D}{4\rho}\partial_j\bm\nabla\cdot(\rho\pvec)\Big]\nonumber\\
&&-\frac{\alpha\tm}{4\rho}\delta_{ij}^T\partial_k\Big\{\rho\Big[\partial_j(\rho\phat_k)+
\partial_k(\rho\phat_j)+\delta_{jk}\bm\nabla\cdot(\rho\pvec)\Big]\Big\}\nonumber\\
&&-\frac{\alpha\tm}{3\rho}\delta_{ij}^T\Big[2\partial_k(\rho\phat_k\partial_j\rho)
+\partial_k(\rho\phat_j\partial_k\rho)+\partial_j(\rho\pvec\cdot\bm\nabla\rho)\Big]
\;,
\end{eqnarray}
where
\begin{equation}
\delta_{ij}^T(\pvec)=\delta_{ij}-\phat_i\phat_j\;,
\end{equation}
projects in the direction transverse to ${\bf \hat{p}}$. The usual
elastic constants $K_1$ and $K_3$ for splay and bend deformations,
respectively, can be identified as $K_3\sim 5D/8$ and
$K_1-K_3\sim D/4$.

The first term on the right hand side of the density equation (Eq.~(\ref{rhoeqn_pol_state})) 
is simply filament diffusion. The second term proportional to
$\alpha$ opposes diffusion and describes the effect of filament
bundling. Finally, the last term describes higher order
nonequilibrium couplings between density and polarization.

The broken directional symmetry of the polarized state yields an
effective drift velocity $\sim\beta\tm\rho\pvec$ describing filament
advection along the direction of polarization. This leads to
convective-type terms on the right hand side of both the density and
polarization equations.  These are true nonequilibrum terms that
cannot be obtained from derivatives of a free energy.  They arise
because, due to the anisotropy of rod diffusion, a motor cluster
cross-linking two filaments can yield a net velocity of the pair, even
in the absence of net forces, as shown in Appendix
\ref{app:microscopic}. This term is absent in descriptions of
the hydrodynamics of active polymer solutions and gels {\em close to
  equilibrium} proposed on phenomenological grounds on the
basis of symmetry argument \cite{Kruse04,Kruse05,Voituriez06}. It
is therefore a {\em far from equilibrium} contribution to active
filament dynamics. Kruse et al. \cite{Kruse04,Kruse05} and
Voituriez et al. \cite{Voituriez06} have considered the
hydrodynamics of an active polymer solution including explicitly
the flow of the solvent. In their formulation activity enters via
a chemical potential proportional to ATP concentration. In our
approach this corresponds to the density $\tm$ of active motors.
The polarization equation considered in \cite{Voituriez06}
contains a term like our $\sim\beta\partial_j\rho$ in
Eq.~(\ref{phateqn_pol_state}) and is obtained there by allowing a
coupling between density and splay deformations in the free energy
of the system. In equilibrium polar fluid this term is ultimately
responsible for the instability of a uniformly polarized phase to
splay deformation~\cite{KungMCMKS06,Blankschtein85}.

A nonequilibrium convective-type term of the form contained on the
left hand side of Eq.~(\ref{phateqn_pol_state}) was included in the
hydrodynamic equations introduced by Simha and Ramaswamy
\cite{ramaswamy02} to describe the dynamics of self propelled
nematic particles in a solution. In that case the effect of
self-propulsion was incorporated by assuming that the particles
have a mean drift in the direction of polarization relative to the
solvent, yielding an advection term of the type obtained here.

In the polarization equation it is apparent that rotational
effects from polar crosslinks ($\gamma_P$) increase the bend
stiffness, but do not renormalize the splay elastic constant.
Nonpolar crosslinks ($\gamma_{NP}$) play a role similar to that of
excluded volume corrections in suppressing rotational diffusion.
This is not surprising as nonpolar crosslinks enhance nematic
order in the system.  Filament bundling described by $\alpha$
renormalizes both the splay and bend stiffness and promotes
spatial inhomogeneities in the polarization.

\subsubsection{Linear Stability}

\begin{figure}
\center \resizebox{0.7\textwidth}{!}{%
  \includegraphics{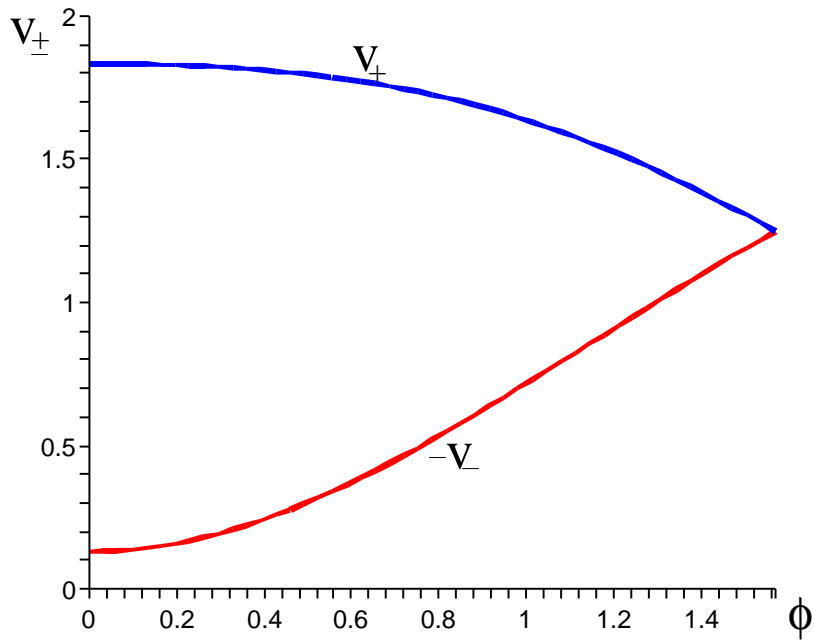}
}

\caption{The propagation speeds $v_+(\phi)$ (blue curve online) and
$-v_-(\phi)$ (red curve online) of the hydrodynamic modes in the
polarized state as a function of the angle $\phi$. The speed
$v_\pm$ is measured in units of $lD_r$ and we have used $\tm=1$,
$\beta=1$ and $\rho_0=\rho_N$.}
\label{v_pm}       
\end{figure}

\begin{figure}
\center \resizebox{0.7\textwidth}{!}{%
  \includegraphics{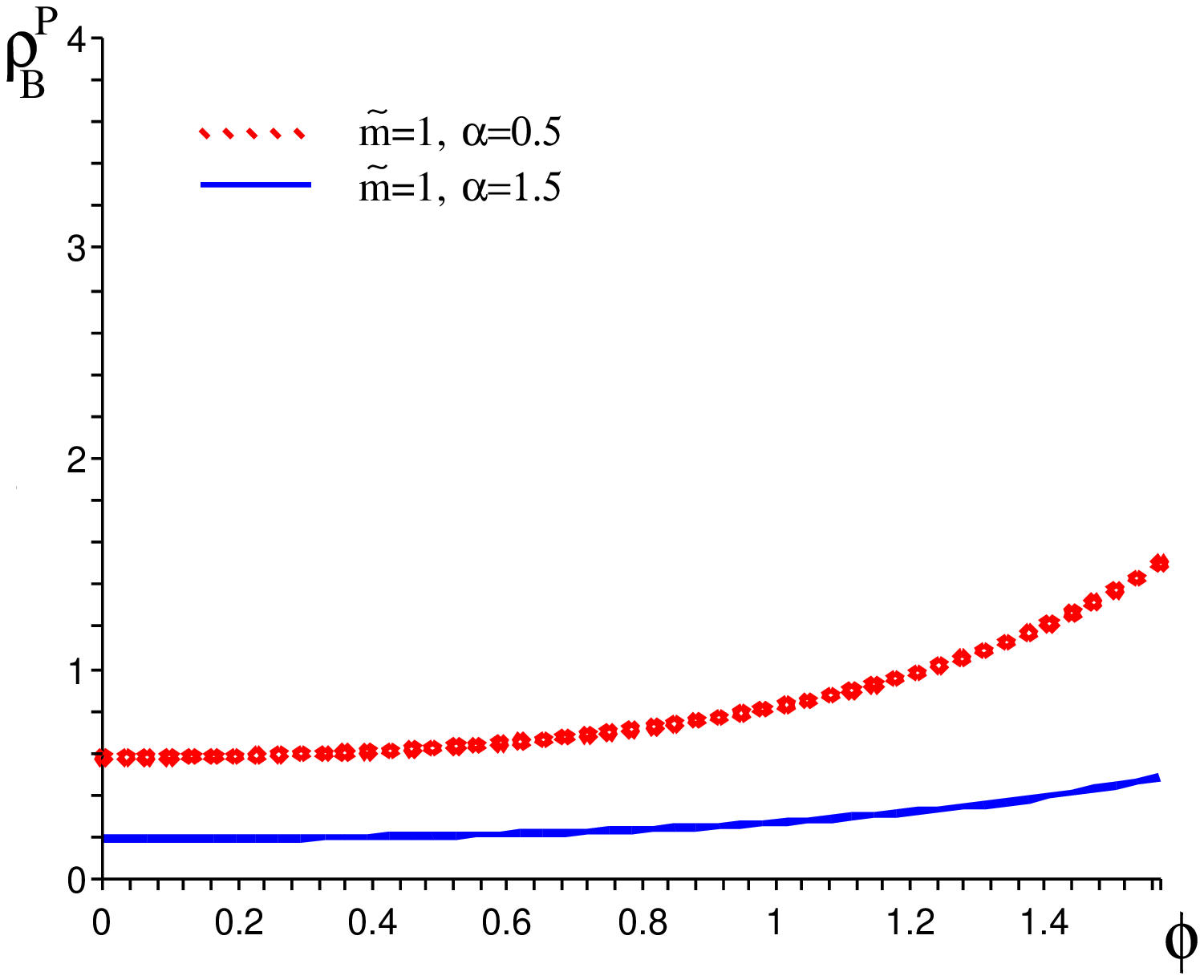}
}

\caption{(color online) The critical density $\rho_B^P$ where the homogeneous
polarized state becomes unstable versus angle $\phi$ for $D=1,~\gamma_{P}=1,~\gamma_{NP}=1$. As in the
nematic phase, by increasing $\alpha$ or $\tm$ the stable region
decreases. The critical density is not very sensitive to changes
in $\gamma_{P}$.}
\label{rho_P}       
\end{figure}

To examine the stability of the polarized state we choose the
$\bf\hat{y}$ axis along the direction of broken symmetry and
expand the hydrodynamic fields about their equilibrium values as
\begin{eqnarray}
&& \rho(\bfr)=\rho_0+\delta\rho(\bfr)\;,\\
&&\pvec(\bfr)={\bf \hat{y}}+\delta\pvec_\perp(\bfr)\;,
\end{eqnarray}
where $\delta\pvec=\xhat\delta\phat_x+{\cal O}\big((\delta
\phat_x)^2\big)$. Expanding the fluctuations in Fourier
components, the linearized equations are given by
\begin{eqnarray}
\partial_t\rho_{\bfk}=-\Big[D_\alpha^P k^2-\alpha\tm\rho_0k_y^2-2iwk_y\Big]\rho_{\bfk}
+\Big[iwk_x+\alpha\tm\rho_0 k_x k_y\Big]\rho_0\phat_{\bfk}\;,
\end{eqnarray}
\begin{eqnarray}
\partial_t \phat_{\bfk}=-\Big[K_\alpha({\bf \hat{k}})k^2+iw'k_y\Big]\phat_{\bfk}
+\Big[iw''k_x-D'_\alpha k_xk_y\Big]\frac{\rho_{\bfk}}{\rho_0}\;,
\end{eqnarray}
where
\begin{eqnarray}
&&w=\frac{7}{36}\tm\beta\rho_0\;,\\
&&w'=\frac{1}{36}\tm\beta\rho_0\;,\\
&&w''=\frac{13}{36}\tm\beta\rho_0\;,\\
&&D_\alpha^P=\frac{3}{4}\Big(D-2\alpha\tm\rho_0\Big)\;,\\
&&D'_\alpha=\frac{1}{4}\Big(D+\frac{\tm}{24}\gamma_{NP}\rho_0-6\alpha\tm\rho_0\Big)\;,\\
&&K_\alpha({\bf
\hat{k}})=\frac{1}{4}\Big(\frac{5}{2}D+\frac{\tm}{6}\gamma_P\rho_0-\tm\alpha\rho_0\Big)\frac{k_y^2}{k^2}
+\frac{1}{4}\Big(\frac{7}{2}D+\frac{\tm}{6}\gamma_P\rho_0-3\tm\alpha\rho_0\Big)\frac{k_x^2}{k^2}\;.
\end{eqnarray}
Note that $K_\alpha({\bf \hat{k}})$, with ${\bf \hat{k}}={\bf
k}/k$, is a generalized stiffness for splay ($k_y=0$) and bend
($k_x=0$) deformations. Denoting by $\phi$ the angle between the
wavevector ${\bf k}$ and the direction of broken symmetry the
hydrodynamic modes describing the decay of density and
polarization fluctuations are given by
\begin{equation}
z_{\pm}(k,\phi)=ikv_\pm(\phi)-{\cal D}^P_\pm(\phi)k^2\;.
\end{equation}
The modes are always propagating with speed
\begin{eqnarray}
v_\pm(\phi)=(w-w'/2)\cos\phi\pm\sqrt{(w+w'/2)^2\cos^2\phi+ww''\sin^2\phi}\;.
\end{eqnarray}
The angular dependence of the speed of propagation is shown in
Fig.~\ref{v_pm}. The decay rate is given by
\begin{eqnarray}
{\cal D}^P_\pm(\phi)&=&\frac{1}{2}\Big[D_\alpha^P+K_\alpha(\phi)-\tm\alpha\rho_0\cos^2\phi\Big]\nonumber\\
&&\pm
\frac{1}{2}\cos\phi\frac{(w+w'/2)\big[D_\alpha^P-K_\alpha(\phi)-\tm\alpha\rho_0\cos^2\phi\big]+\sin^2\phi(wD'_\alpha-w''\tm\alpha\rho_0)}
{\sqrt{(w+w'/2)^2\cos^2\phi+ww''\sin^2\phi}}\;.
\end{eqnarray}
For large values of the bundling rate $\alpha$ the various elastic
constants are driven to zero and  ${\cal D}^P_\pm(\phi)<0$,
corresponding to the instability of the uniform polarized state.
The condition ${\cal D}^P_\pm(\phi)=0$ defines the value
$\rho_B(\phi)$ of the filament density above which the polarized
state is unstable. This value is largest for $\phi=\pi/2$,
corresponding to fluctuations with ${\bf k}$ normal to the
direction of mean polarization (i.e., pure splay deformations of
the local polarization) due to the stiffening of the splay elastic
constant from polar crosslinks. In contrast, the bend stiffness is
not renormalized by polar crosslinks, resulting in a lower value
of $\rho_B$ at $\phi=0$, where polarization deformations are pure
bend. The angular dependence of $\rho_B(\phi)$ is shown in
Fig.~\ref{rho_P}.

Finally, it is useful to consider explicitly the two limiting
cases $\phi=0$ and $\phi=\pi/2$. For $\phi=0$ (i.e., $k_y=k$)
density and polarization (in this case bend deformations)
fluctuations are decoupled. Their respective relaxation rates are
given by
 \begin{eqnarray}
 &&z_+(k,\phi)\equiv z_\rho(k,\phi)=2iwk-\Big[\frac{3}{4}D-\frac{5}{2}\tm\alpha\rho_0\Big]k^2\;,\\
 &&z_-(k,\phi)\equiv z_p(k,\phi)=-iw'k-\frac{1}{4}\Big[\frac{5}{2}D+\frac{\tm}{6}\gamma_P\rho_0-\tm\alpha\rho_0\Big]k^2\;.
 \end{eqnarray}
The bundling instability is controlled by the growth of density
fluctuations and occurs at
\begin{equation}
\rho_B(\phi=0)=\frac{3D}{10\tm\alpha}\;.
\end{equation}

For $\phi=\pi/2$ (i.e., $k_x=k$) the modes are complex conjugate,
with
\begin{equation}
z_\pm(k,\phi=\pi/2)=\pm
ik\sqrt{ww''}-\frac{k^2}{8}(13D/2+\tm\gamma_{P}\rho_0/6-9\tm\alpha\rho_0)\;.
\end{equation}
Both density and splay fluctuations of the polarization field go
unstable at the same density, given by
\begin{equation}
\rho_B(\phi=\pi/2)=\frac{13D}{\tm(18\alpha-\gamma_P/3)}\;.
\end{equation}
The zipping effect described by $\gamma_P$ tends to stabilize the
system.

\subsection{Summary}
\label{subsec:summary} 
\begin{figure}
\center \resizebox{0.7\textwidth}{!}{%
 \includegraphics{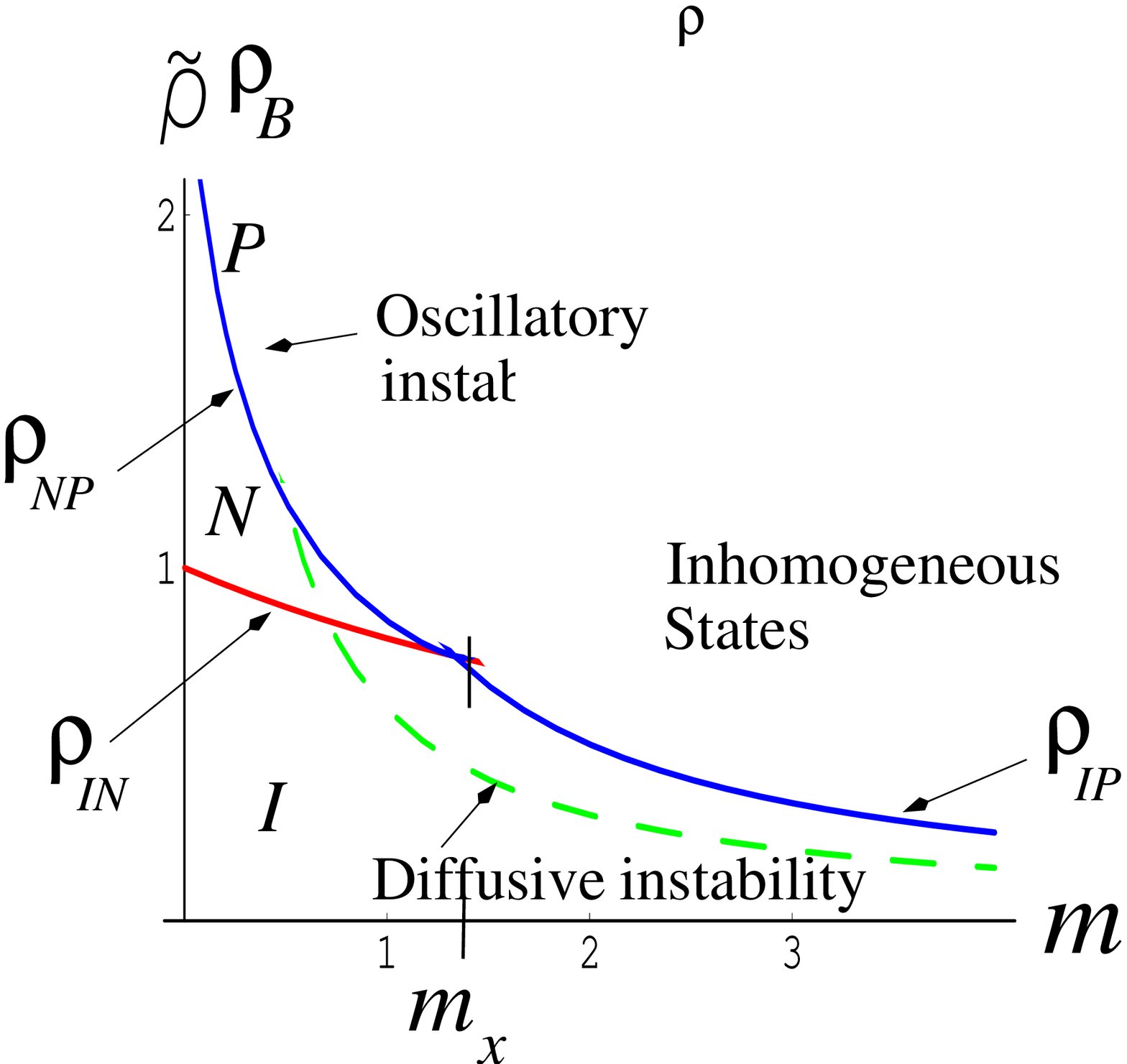}
 }

\caption{Bundling renders homogeneous states unstable for
$\rho_0>\rho_B$, where they are replaced by inhomogeneous
solutions. The $\rho_B$ line may lie above the $\rho_{NP}-\rho_{IP}$ line
or cross through the {\em N} and {\em I} states, as shown in the figure
($\gamma_P/\gamma_{NP}=1$, $c_2=50$, $\gamma_{NP}/\alpha=0.6$), depending on
the value of $\gamma_{NP}/\alpha$. The instability of the {\em I} and {\em N} states
is diffusive (dashed line), while the instability of the {\em P} state is oscillatory (dotted line).}
\label{rho_instability}
\end{figure}

All homogeneous states are rendered
unstable by the same mechanism of filament bundling, driven by the
parameter $\alpha$. Up to numerical constants and assuming
$\gamma_P<\alpha$, the density above which the homogeneous states
are unstable can be estimated as $\rho_B\sim D/\tm\alpha$. The
bundling instability line is shown in Fig.~\ref{rho_instability}.
One important observation is that the nature of the instability
changes from diffusive in both the isotropic and the nematic
states to oscillatory in the polarized state. This suggests that
at high filament and motor density the uniform polarized state may
be replaced by spatially inhomogeneous oscillatory structures such
as vortices.

\section{Discussion}
Several other authors have recently put forward descriptions of
the dynamics of active solutions and gels of long biofilaments and
molecular motors. It is useful to compare our work to others in
some detail.

Kruse et al. \cite{Kruse04,Kruse05} and Voituriez et al.
\cite{Voituriez06} have developed a continuum phenomenological
descriptions of the polarized state of active polymer solutions
where the hydrodynamic equations are written down on the basis of
general symmetry considerations. Our work, in contrast, derives
such equations from a systematic coarse-graining of a more
microscopic kinetic equation. The advantage of the former method
is its generality. The disadvantage is that the resulting
continuum equations contain many undetermined parameters. Our work
yields an estimate for these parameters and an understanding for
the microscopic mechanisms that control each term in the continuum
equations. On the other hand, the precise dependence of the
parameters on the physical properties of the crosslinkers is
determined by the specific microscopic model considered, as shown
in Appendix \ref{app:microscopic}. The two approaches are clearly
complementary and both provide insight in the system's dynamics.

The work described in Refs. \cite{Kruse04,Kruse05} and
\cite{Voituriez06} explicitly incorporates the dynamics of the
solvent, which is assumed quiescent in our work (see, however, Ref.
\cite{TBLMCMrheo}), but consider systems near equilibrium by only
keeping terms of first order in the chemical potential which
controls the rate of ATP consumption. For a more precise
comparison we refer to Ref.~\cite{Voituriez06}, where the
equations are written in the simpler form appropriate to an active
viscous solution, with no viscoelastic effects. In
Ref.~\cite{Voituriez06} activity is controlled by the difference
$\Delta\mu$ in chemical potential of ATP and its hydrolysis
products, assumed to be constant. This corresponds in our work to
the product $\tm u_0=\tm a R_{ATP}$ which controls ATP consumption
in the system. All active contributions are proportional to the
combinations $\tm R_{ATP}$. The parameter $\lambda$ of
Ref.~\cite{Voituriez06} corresponds to our polar rotational rate
$\gamma_P$, describing the "zipping" of filaments due to the
action of polar crosslinkers and responsible for establishing the
homogeneous polarized state. This term is ignored in
Ref.~\cite{Voituriez06}, where it is assumed from the outset that
the system exists in a polarized state with $p_0=1$. The terms
proportional to the parameter $w$ arising in the polarization
equation of Ref.~\cite{Voituriez06} have the same structure as the
first term on the right hand side of our polarization equation,
Eq.~(\ref{phateqn_pol_state}). However, this term is obtained in
Ref.~\cite{Voituriez06} as derivative of a phenomenological free
energy, consisting of the usual Frank free energy for a nematic
plus a term $\sim w\rho\bm\nabla\cdot{\bf p}$, which is allowed in
a polar fluid. As a result, this term does not appear explicitly
as an active term proportional to $\Delta\mu$. On the other hand,
our analysis of the homogeneous states shows that activity is
probably crucial for establishing the polarized state as the
zipping rate $\gamma_P$ induced by polar crosslinkers is likely to
depend on ATP consumption rate. Furthermore, the convective terms
on the left hand side of Eqs.~(\ref{rhoeqn_pol_state}) and
(\ref{phateqn_pol_state}) cannot be obtained as derivatives of a
free energy and are therefore not present in
Ref.~\cite{Voituriez06}. These terms are linear in the ATP
consumption rate and are important in controlling the oscillatory
nature of the spatial patterns (e.g., vortices) that are obtained
at high filament and motor density \cite{Aranson05}.

Aranson and Tsimring \cite{Aranson05} have used a generalization
of the Maxwell model of binary collisions in a gas to describe the
dynamics of a solution of polar rods with inelastic interaction
representing the effect of active crosslinkers. Although their
kinetic model, in contrast to ours, allows for instantaneous large
changes in the relative angle of two rods upon collision, the
continuum equations for density and polarization obtained from the
model have the same structure as ours. Our parameter $\alpha$
corresponds essentially to their parameter $B^2$ (related to the
spatial range of the interaction between two rods), while our
parameter $\beta$ is proportional to their parameter $H$, which
controls the strength of the dependence of the interaction between
two rods on their relative orientation (although again these
authors do not include the convective terms $\sim\beta$ in the
density and polarization equations). The dependence on motor
density or ATP consumption rate does not appear explicitly in the
continuum equations of Ref.~\cite{Aranson05} as the strength of
the motor-mediated interactions is scaled out of the calculation.

One important difference between our work and that of
Ref.~\cite{Aranson05} is that by incorporating excluded volume
effects and including the action of both stationary and mobile
crosslinkers, we can obtain a complete characterization of the
 homogeneous states of the system. In particular, we show that
 both nematic and polar order are possible in different regions of
 parameters, and evaluate the effect of crosslinks on the
 isotropic-nematic transition.
 
 Our work can be extended in several ways. First, we have assumed that
 the solvent is quiescent and only provides the damping on the
 dynamics of filaments and motors. Relaxing this approximation
 requires considering explicitly the dynamics of a two-component
 system. In particular the dynamics of the solvent must be
 incorporated when considering the response of the system to an
 externally imposed flow. This will be discussed in a future
 publication \cite{TBLMCMrheo}. Secondly, an analysis of the nonlinear
 equations for the director and polarization fields in the nematic and
 polarized phases, respectively, reveals the structure of the possible
 topological defects in each phase and their stability. This analysis
 can be carried out partly analytically and partly numerically and can
 be used to study the range of stability of the spatially
 inhomogeneous patterns seen in the in vitro experiments by
 considering each pattern as composed of topological defects of the
 bulk system. Finally, for comparison with experimental systems it is
 crucial to consider the dynamics of active solutions in specific
 geometries, with suitable boundary conditions \cite{Claessens}. An
 important application of the dynamics of active filament solutions
 and gels is that of cell locomotion on a substrate. This may be
 modeled by considering a thin active layer on a substrate, but will
 require incorporating in the model the nonequilibrium
 polymerization-depolymerization of the filaments, mechanical coupling to the
 substrate and understanding the interplay between them and activity.

\acknowledgments We thank Sriram Ramaswamy for useful discussions.
This work was supported by the Royal Society (TBL) and  by the
National Science Foundation, grants DMR-0305407 and DMR-0219292
(AA and MCM).

\appendix

\section{Microscopic models of motor-filament kinematic}
\label{app:microscopic}

In this Appendix we describe some microscopic models of the
motor-mediated interaction among two filaments.  Clearly such models
are a great simplification of the contributions to the motor-mediated
forces, but they allow us to estimate the various phenomenological
parameters introduced in Section~\ref{sec:Model} and to justify the
approximations used in this paper.  We consider two classes of models:
1) small motor clusters with an inhomogeneous stepping velocity
that vanishes at the plus end of the filament, inspired by kinesin
motor constructs interacting with microtubules; 2) filamentous motor
clusters with an antiparallel arrangement of heads inspired by thick myosin
filaments interacting with thin actin filaments.

\subsection{Stalling clusters}
\label{sub:stalling}
 The model
presented here extends the one discussed in Ref.~\cite{TBLMCM03}
to include the possibility of motor-induced filament rotation.


We consider a pair of filaments (denoted as filaments 1 and 2)
cross-linked by a motor cluster.  Due to the action of the motors,
filaments 1 and 2 acquire center-of-mass velocities ${\bf v}_1$
and ${\bf v}_2$ and rotational velocities $\bm\omega_1$ and
$\bm\omega_2$ about the center of mass. Our goal is to evaluate
these velocities in terms of the rates at which the motor cluster
steps along the filament and rotates relative to it, and of the
filaments' orientation. Both filaments and motors move through a
solution. We assume that the filament dynamics is overdamped and
the friction of motors is very small compared to that of
filaments. The coupling of the filaments to the fluid is via a
local friction (Rouse model). This is a reasonable approximation
for a quiescent solution without externally imposed flow nor flow
generated by the motor activity. Under these conditions,
hydrodynamic coupling yields logarithmic corrections to the
friction, which are small for long thin
rods~\cite{TBL_active,TBL_active2,DoiEdwards}. Momentum conservation
requires that in the absence of external forces and torques, the
total force (torque) acting on filaments centered at a given
position be balanced by the frictional force (torque) experienced
by the filament while moving through the fluid. The third law
requires that any force or torque generated by an active crosslink
on one of the filaments of the pair is balanced by an equal and
opposite force or torque acting on the other filament. This yields
\begin{eqnarray}
\label{force_bal} & &\zeta_{ij}(\nvec_1)v_{1j}= - \zeta_{ij}(\nvec_2)v_{2j}\;,\\
& & \label{torque_bal}  \zeta_r\bm{\omega}_1=-
\zeta_{r}\bm{\omega}_2\;,
\end{eqnarray}
where
$\zeta_{ij}(\nvec)=\zeta_\parallel\nhat_i\nhat_j+\zeta_\perp(\delta_{ij}-\nhat_i\nhat_j)$
is the friction tensor of the rod, with $\zeta_\parallel$ and
$\zeta_\perp$ the longitudinal and transverse friction
coefficients, respectively, and $\zeta_r$ is the rotational
friction. Equation~(\ref{force_bal}) shows that the anisotropy of the
friction tensor allows for a net translation (${\bf v}_1+{\bf
  v}_2\not=0$) of the filament pair induced by motors.

\begin{figure}
\center \resizebox{0.5\textwidth}{!}{%
  \includegraphics{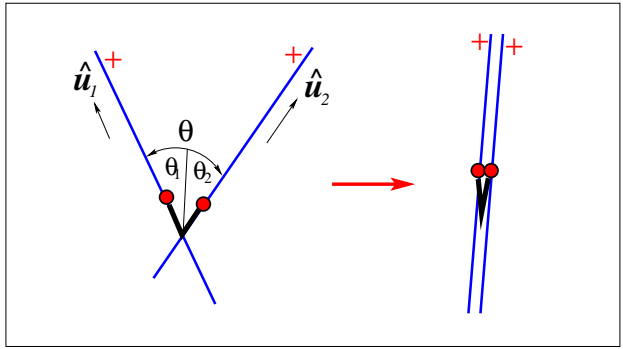}
}

\caption{
  Two filaments of orientation $\nvec_1$ and
  $\nvec_2$ connected by an active torsional spring. Here $\theta_i=\theta(s_i)$
, for $i=1,2$ is the torsional angle at the point of
attachment of the motor cluster. Note that we have chosen a
convention such that $\theta=\theta_1-\theta_2$.}
\label{2filam_torque}       
\end{figure}

The mobile crosslink is a small aggregate of molecular motors that
exerts forces and torques on the filaments by converting chemical
energy from the hydrolysis of ATP into mechanical work.  While
walking along the filaments, the motor clusters can also apply
aligning torques on the filaments, if there is a preferred angle
between the heads of the motor cluster. To capture this, we
consider the cluster to be a nonlinear torsional spring of size
$l_m\sim b<<l$. A similar description would also be appropriate for any
polar cross-linking protein.  However, a motor cluster which
aligns the filaments by active contractions has an ATP-dependent
spring constant.  A schematic is shown in Fig.~\ref{2filam_torque}.

\begin{figure}
\center \resizebox{0.5\textwidth}{!}{%
  \includegraphics{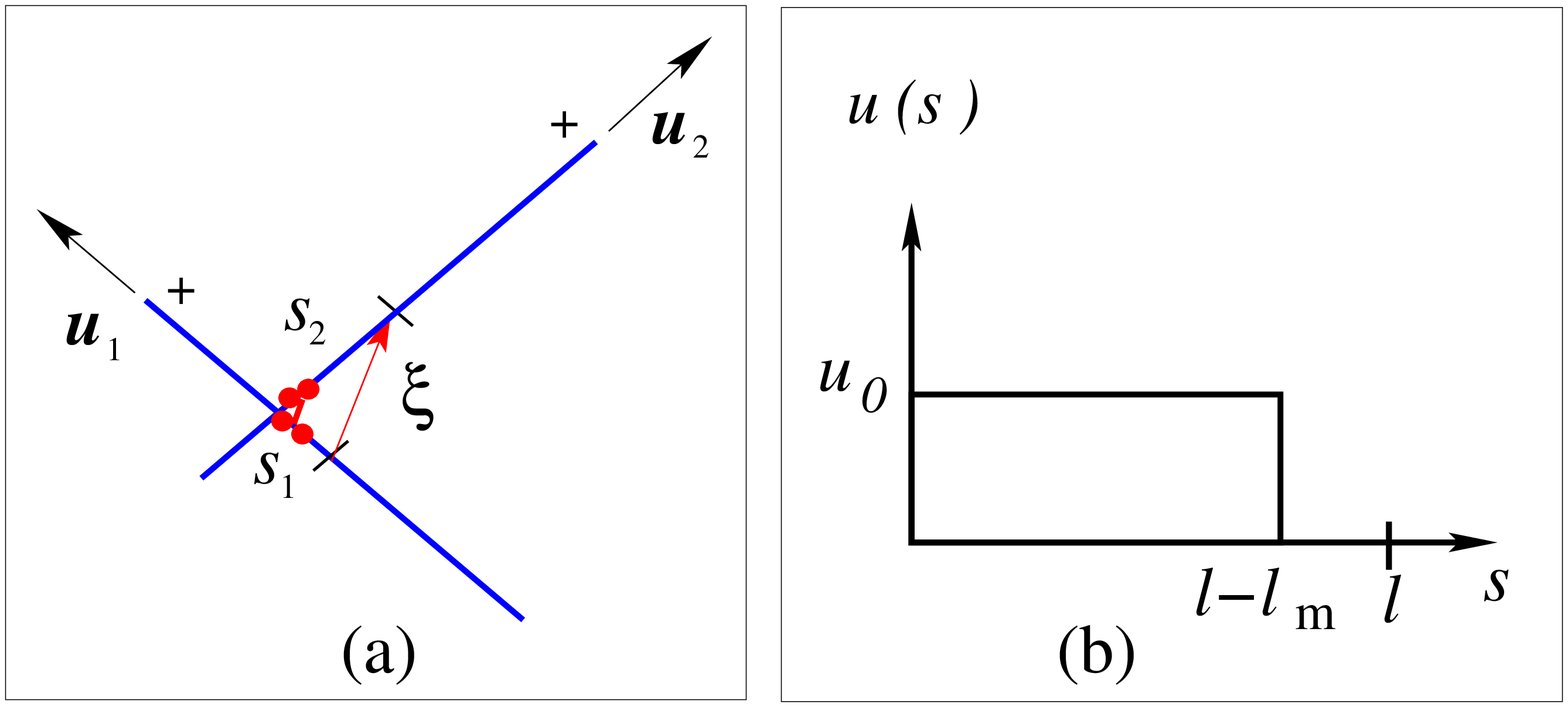}
}

\caption{(a) Two filaments connected by an active cross-link
and the geometry of the overlap. The filaments' centers are
separated by $\bxi=s_1\nvec_1 - s_2\nvec_2$. (b) The profile of
the motor stepping rate.}
\label{2filam}       
\end{figure}

It is convenient to think of the motor cluster as composed of two
heads, with the i-th head ($i=1,2$) attached to the
i-th filament at position ${\bf r}_i^{\times}={\bf
  r}_i+\nvec_i s_i$. Motor heads are assumed to step
towards the polar end of filaments at a known speed, $u(s)$, which
depends on the point of attachment (see Fig.~\ref{2filam}). The
motor-induced torques occur along the axis of the motor cluster,
assumed to be directed perpendicular to the plane containing the
filaments, and are capable of generating equal and opposite torques
on the two filaments. The torsional angles $\theta(s)$ obey the following
equations
 $\zeta_r \dot{\theta}(s_1)=-\zeta_r
\dot{\theta}(s_2)=-\kappa \sin\big[\theta(s_1)-\theta(s_2)\big]
\simeq -\kappa \big[\theta(s_1)-\theta(s_2)\big]$,
$\kappa$ is the torsional spring constant.  In general the
torsional spring constant will also depend on the position of
the motor cluster along the filament, i.e., $\kappa=\kappa (s)$.
The resulting inhomogeneities in the rotational rate $\gamma$ does
not yield qualitatively new terms in the hydrodynamic equations and
will be neglected here (see also~\cite{nonlinear_polarized}). The dynamics of the
i-th motor head is described by a translational
velocity ${\bf v}^m_i$ at the point of attachment and a
rotational velocity $\bm\omega^m_i$, given by
\begin{eqnarray}
\label{motor_vel} &&{\bf v}^m_i =\dot{\bf
r}_i^{\times}={\bf
v}_i+u(s_i)\nvec_i+s_i\bm{\omega}_i\times\nvec_i\;,\\
\label{motor_omega}&&\bm{\omega}^m_i=\bm\omega_i
+(-1)^{i-1}{\kappa \over \zeta_r }~\frac{\nvec_1\times\nvec_2}{|\nvec_1\times\nvec_2|}\;,
\end{eqnarray}
with $u(s_i)=\dot{s}_i$ . Finally,  the two motors
within a cluster are rigidly attached to each other. This requires
\begin{eqnarray}
& & {\bf v}^m_1={\bf v}^m_2\;,\label{motorv}\\
& &\bm{\omega}^m_{1}=\bm{\omega}^m_{2}\;.\label{motorrot}
\end{eqnarray}

Since the motor cluster has size $l_m\sim b<<l$, we can neglect
the length of the motor compared to that of the filament and
assume that the attachment points satisfy ${\bf
r}_1^{\times}={\bf r}_2^{\times}$, or $\bm{\xi}={\bf
r}_2-{\bf r}_1= s_1\nvec_1-s_2\nvec_2$. Equations
(\ref{force_bal}) and (\ref{torque_bal}), together with the
expressions (\ref{motor_vel}) and (\ref{motor_omega}) for the
velocities of the motor heads, and the conditions (\ref{motorv})
and (\ref{motorrot}) that the two motor heads are rigidly
connected, then provide a closed set of equations that can be
solved to obtain the filaments' translational and rotational
velocities in terms of their relative orientations and of the
motors' stepping and torsion rates.  It is convenient to introduce
relative and net translational and rotational velocities of the
filament pair as
\begin{eqnarray}
&&{\bf v}={\bf v}_1-{\bf v}_2\;,\\
&&{\bf V}=\frac{{\bf v}_1+{\bf v}_2}{2}\;, \end{eqnarray}
and
\begin{eqnarray}
&&\bm\omega=\bm\omega_1-\bm\omega_2\;,\\
&&\bm\Omega=\frac{\bm\omega_1+\bm\omega_2}{2}\;. \end{eqnarray}
For the translational velocities we obtain
\begin{eqnarray}
\label{vel_eq2} {\bf v}&=&u(s_2)\nvec_2-u(s_1)\nvec_1+\gamma(s_2-s_1)(\nvec_1+\nvec_2)
+\gamma(1+\nvec_1\cdot\nvec_2)(s_1\nvec_1-s_2\nvec_2)\nonumber\\
&=&\frac{1}{2}\Big\{[u(s_2)-u(s_1)]-\gamma(s_1-s_2)(1-\nvec_1\cdot\nvec_2)\Big\}(\nvec_2+\nvec_1)\nonumber\\
&&+\frac{1}{2}\Big\{[u(s_2)+u(s_1)]-\gamma(s_1+s_2)(1+\nvec_1\cdot\nvec_2)\Big\}(\nvec_2-\nvec_1)\;,
\end{eqnarray}
\begin{eqnarray}
\label{V} {\bf V}&=&A(\nvec_2+\nvec_1)+B(\nvec_2-\nvec_1)\;,
\end{eqnarray}
with
\begin{eqnarray}
\label{Vnet_app} A&=&-\frac{\sigma}{4}\Big(\frac{1-\nvec_1\cdot\nvec_2}{1-\sigma\nvec_1\cdot\nvec_2}\Big)
\Big[u(s_2)+u(s_1)-\gamma(s_1+s_2)(1+\nvec_1\cdot\nvec_2)\Big]\;,\\
B&=&\frac{\sigma}{4}\Big(\frac{1+\nvec_1\cdot\nvec_2}{1+\sigma\nvec_1\cdot\nvec_2}\Big)
\Big[u(s_2)-u(s_1)+\gamma(s_2-s_1)(1-\nvec_1\cdot\nvec_2)\Big]\;,
\end{eqnarray}
where we have defined $\gamma=\kappa/\zeta_r$ and $\displaystyle \sigma={(\zeta_\perp-\zeta_\parallel) \over
(\zeta_\perp+\zeta_\|)}$. For long thin rods
$\zeta_\perp=2\zeta_\parallel\equiv 2\zeta$ and $\sigma=1/3$.

There is no net rotational velocity of the pair ($\bm\Omega=0$).
The relative rotational velocity is given by

\begin{equation}
\label{omega_micro} \bm\omega=2 \gamma ~\frac{\nvec_1\times\nvec_2} {|\nvec_1\times\nvec_2|}\;.
\end{equation}
The fact that ${\bf V}\not=0$ indicates that motor activity can
induce a net motion of the pair relative to the solution. This is
a consequence of hydrodynamics at low Reynolds numbers, which
gives an anisotropic of friction tensor for long thin rods.  As a
result ${\bf V}$ vanishes when $\zeta_\perp=\zeta_\parallel$. Also ${\bf V}$
vanishes identically for $\nvec_2=\pm\nvec_1$, so that ${\bf V}=0$
in one dimension.

We can compare the expression for the filament velocities obtained
via the microscopic model described in this section to the general
expression introduced on the basis of symmetry considerations in
Eqs. (\ref{vrel}-\ref{omega}) by expanding the stepping rate as
$u(s)\approx u_0-su'$, where $u'=-du(s)/ds>0$. Substituting the expressions
$\frac{1}{2}[u(s_1)-u(s_2)]\simeq\frac{u'}{2}(s_1-s_2)$ and
$\frac{1}{2}[u(s_2)+u(s_1)\simeq u_0+\frac{u'}{2}(s_2+s_1)$
into Eq. (\ref{vel_eq2}) we obtain a general expression for the relative velocity given by
\begin{eqnarray}
\label{vrel_app}
\bf{v}&=&\alpha_{+}\nvec_{+}(\bxi\cdot\nvec_{+})+\alpha_{-}\nvec_{-}(\bxi\cdot\nvec_{-})
+\beta(\nvec_2-\nvec_1)\;,
\end{eqnarray}
where $\nvec_{+}=(\nvec_2+\nvec_1)/|\nvec_2+\nvec_1|$ and  $\nvec_{-}=(\nvec_2-\nvec_1)/|\nvec_2-\nvec_1|$ and
\begin{eqnarray}
\alpha_{+}&=&-\gamma(1-\nvec_1\cdot\nvec_2)+u'\;,\\
\alpha_{-}&=&\gamma(1+\nvec_1\cdot\nvec_2)+u'\;,\\
\beta &=& u_0\;.
\end{eqnarray}
If $u' \gg \gamma$, then
$\alpha_{+}=\alpha_{-}=\talpha$, leading to the simpler expression for the
relative velocity (see Eqs. (\ref{vrel}-\ref{omega})) which we use for
the whole of this article .

By comparing Eqs.~(\ref{vrel_app}),
(\ref{Vnet_app}) and (\ref{omega_micro}) to the general expressions
given in Sec.  \ref{sec:Model} we obtain the following estimates
\begin{eqnarray}
&&\talpha_0\simeq  l |\frac{du}{ds}|\;,\\
&& \beta_0\simeq u_0\;,\\
&& \gamma_P\sim \kappa / \zeta_r \;.
\end{eqnarray}
Note that the specific microscopic model used here gives $\talpha_1
=0,\beta_1=0$ and $\gamma_{NP}=0$. This is the result of having
considered the kinematics of a single pair of filaments coupled by one
motor cluster. An additional dependence of the effective coupling
constants is introduced by the dependence of the motor binding
probability on the relative orientation of the
filaments.

\subsection{Contractile motor filaments}

Here we consider another microscopic model relevant to large
contractile filaments of myosin II (thick mini-filaments) interacting
with filamentous actin (thin filaments).  Both the thick contractile
motor filament and the thin filaments undergo overdamped motion in a
quiescent fluid.

\begin{figure}
\center \resizebox{0.7\textwidth}{!}{%
  \includegraphics{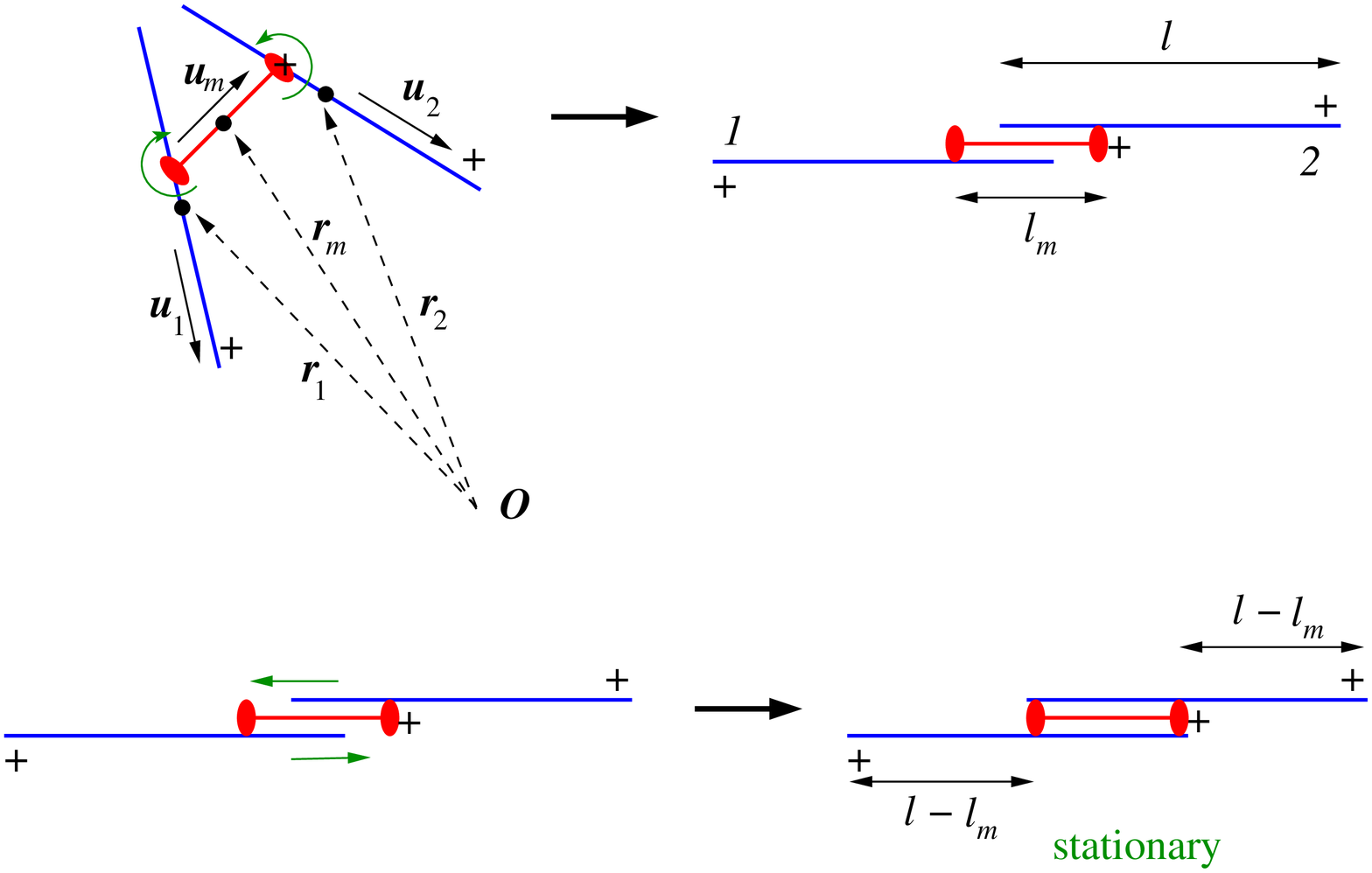}
}

\caption{
  Two thin filaments of length $l$ connected by a
  contractile thick motor filament of length $l_m$. Due to the
  torsional springs, the thick filament aligns the actin filaments in
  an antiparallel configuration. The 'stepping' of motor heads towards
  the plus ends of the thin filaments bring the (almost) antiparallel
  filaments together if their centres are more than $l-l_m$ apart.
  When their centers of mass are separated by $l-l_m$, the two actin
  filaments are stationary due to the opposing effects of the two
  motor heads at the opposite ends of the motor filament.  }
\label{fig:motor-fil}       
\end{figure}

We consider two thin (e.g., actin) polar filaments of length $l$ with
centers of mass at ${\bf r}_1$ and ${\bf r}_2$ and orientations
$\nvec_1$ and $\nvec_2$, respectively. The active crosslink is a
(thick) filament of motor proteins of length $l_m < l$.  The motor
heads within the motor filament are antialigned, with the motor heads
at the two ends of the motor filament pointing in opposite directions.
Its orientation is described by a unit vector $\nvec_m$ oriented along
its axis. We choose the direction of $\nvec_m$ to be from thin
filament 1 to 2, as indicated in Fig.~\ref{fig:motor-fil}. The motor
filament exerts torques on the actin filaments by acting as a
torsional spring of strength $\kappa$. As a result, the actin
filaments align with the motor filament, as shown in
Fig.~\ref{fig:motor-fil}.  Once the thin filaments are parallel to the
motor filament, the heads on either side of the motor filament pull
the thin filaments together until they overlap over a length $l_m<l$.
In this configuration both thin filaments are linked by both heads at
the two ends of the contractile motor filament. The effect of these
two heads balance and the thin filaments remain stationary relative to
each other.

To describe the dynamics, we denote by ${\bf v}_m$ and $
{\bm\omega}_m$ the center of mass and angular velocity of the
thick motor filament, respectively.  The friction tensor of the
motor filament is given by
$\zeta^m_{ij}=\zeta_{m\parallel}\nhat_{mi}\nhat_{mj}+\zeta_{m\perp}(\delta_{ij}-\nhat_{mi}\nhat_{mj})$.
Since the thick motor filament is shorter than the two thin
filaments it crosslinks, we expect the motor filament
translational friction coefficients $\zeta_{m\parallel}$ and
$\zeta_{m\perp}$ and rotational friction $\zeta_{mr}$, to be
smaller than the corresponding parameters for the actin filaments,
i.e., $\zeta_{m\parallel},\zeta_{\parallel}$,
$\zeta_{m\perp}<\zeta_{\perp}$, and $\zeta_{mr}< \zeta_{r}$. The
separation of centers of mass of the thin filaments is $\bm{\xi}
\equiv {\bf r}_2 - {\bf
  r}_1 = \nvec_1 s_1 - \nvec_2 s_2 + \nvec_m l_m$.
Force and torque balance require
\begin{eqnarray}
\label{fil_force_bal} &&{\zeta}_{ij}(\nvec_1){v}_{1j}+
{\zeta}_{ij}(\nvec_2){v}_{2j}
+ {\zeta}^m_{ij}(\nvec_m) {v}_{mj}= 0 \;,\\
&& \zeta_r{\bm \omega}_1= {\bf G}_1\;,\nonumber\\
&& \zeta_{r}{\bm \omega}_2 = {\bf G}_2 \;,\nonumber\\
\label{fil_torque_bal}  &&\zeta_{mr} {\bm \omega}_m= -{\bf G}_1 -
{\bf G}_2\;,
\end{eqnarray}
where
\begin{eqnarray}
&& {\bf G}_1 = {\kappa } (\nvec_m \times \nvec_1) \; ,\nonumber\\
&&{\bf G}_2 = - {\kappa } (\nvec_m \times \nvec_2) \;.
\end{eqnarray}
The position of the center of mass of the thick motor filament is
\begin{math}
{\bf r}_m = {\bf r}_1 + s_1 \nvec_1 + \nvec_m l_m/2 = {\bf r}_2 +
s_2 \nvec_2 -\nvec_m l_m/2
\end{math} and its velocity is given by
\begin{eqnarray}
\label{motor_fil_vel} &&{\bf v}^m_i =\dot{\bf r}_m={\bf
v}_i+u(s_i)\nvec_i+s_i\bm{\omega}_i\times\nvec_i
+ (-1)^{i-1} {l_m \over 2}\bm{\omega}_m\times\nvec_m \;, \;
i=1,2\;.
\end{eqnarray}

The set of Eqs.~(\ref{motor_fil_vel}), (\ref{fil_force_bal}) and
(\ref{fil_torque_bal}) can be solved for the mean ${\bf V} = ({\bf
v}_1 + {\bf v}_2)/2\; , \; {\bm \Omega}=\left(
  {\bm \omega}_1 + {\bm \omega}_2 \right)/2$ and relative ${\bf v} =
{\bf v}_1 - {\bf v}_2, \; {\bm \omega}={\bm \omega}_1 - {\bm
    \omega}_2$ translational and angular velocities of the two thin
filaments and the velocity ${\bf v}_m$ of the thick motor
filament. The general solution is complicated and not terribly
transparent.

To simplify the solution, we assume that $l_m \ll l$ so that
$\zeta_{mr}\ll \zeta_r$ and $\zeta_m \ll \zeta$. In this case the
orientation of the thick motor filament relaxes much faster than
that of the two thin filaments and is therefore slaved to the thin
filament orientations, so that $\nvec_m = (\nvec_2 -
\nvec_1)/|\nvec_2 - \nvec_1|$. The expressions for the relative
and mean velocities are then
\begin{eqnarray}
\label{vrel_fil_app}{\bf v}=u(s_2)\nvec_2-u(s_1)\nvec_1 +&&
\frac{\gamma}{2}
\Big\{\frac{\left(1+\nvec_2\cdot\nvec_1\right)}{\left|\nvec_2-\nvec_1\right|}
\left( \nvec_2 -\nvec_1\right) \left( s_2 + s_1\right)\nonumber\\
&& + \frac{\left( 1- \nvec_1 \cdot \nvec_2\right)}{\left| \nvec_2
-\nvec_1\right|} \left(s_1 - s_2 \right)\left( \nvec_2
+\nvec_1\right) \Big\}\;,
\end{eqnarray}
\begin{eqnarray}
{\bf V} =A \left(\nvec_2 + \nvec_1 \right) + B \left(\nvec_2 -
\nvec_1 \right) \label{Vnet_fil_app}\;, \end{eqnarray}
where
\begin{eqnarray}
A &=& - \left({\sigma \over 4} \right) \left({1 - \nvec_1 \cdot
\nvec_2 \over 1 - \sigma \nvec_1 \cdot \nvec_2 }\right)
\left[u(s_2) + u(s_1) + {\gamma}
 {\left( 1 + \nvec_2 \cdot \nvec_1 \right) \over \left| \nvec_2 -\nvec_1\right|} \left( s_2 + s_1\right)
\right]\;, \\
B &=&  \left({\sigma \over 4} \right) \left( {1 + \nvec_1 \cdot
\nvec_2 \over 1 +  \sigma \nvec_1 \cdot \nvec_2
}\right)\left[u(s_2) - u(s_1) + {\gamma} {\left( 1- \nvec_1 \cdot
\nvec_2\right) \over \left| \nvec_2 -\nvec_1\right|} \left(s_1 -
s_2 \right) \right]\;.
\end{eqnarray}

The center of mass velocity of the thick motor filament is given
by
\begin{equation}
{\bf v}_m = {\bf V}+ {\bf w}/2 \;,
\end{equation}
where
\begin{eqnarray}
&& \nonumber \\
{\bf w} =u(s_2)\nvec_2 + u(s_1)\nvec_1 -
&&{\gamma \over 2 } \Big\{ {\left(1+ \nvec_2 \cdot \nvec_1 \right) \over \left| \nvec_2 -\nvec_1\right|}\left( \nvec_2 -\nvec_1\right) \left( s_1 - s_2\right)\nonumber\\
&& + {\left( 1- \nvec_1 \cdot \nvec_2\right) \over \left| \nvec_2
-\nvec_1\right|} \left(s_1 + s_2 \right)\left( \nvec_2
+\nvec_1\right) \Big\}\;,
\end{eqnarray}
with $\displaystyle \sigma={(\zeta_\perp-\zeta_\parallel) \over
(\zeta_\perp+\zeta_\|)}$. There is no net rotational velocity of
the pair ($\bm\Omega=0$) and the relative rotational velocity is
given by
\begin{equation}
\label{omega_micro_fil} \bm\omega=2 \gamma
~\frac{\nvec_2\times\nvec_1} {|\nvec_1\times\nvec_2|}\;.
\end{equation}
where $\gamma = {\kappa \over \zeta_r}$.

The speed of the motor clusters depends on the thin filaments'
relative position and orientation: it is maximal when the
filaments are oppositely oriented and goes to zero when the
filament overlaps a length $l_m$ (see Fig.~\ref{fig:motor-fil}). A
simple expression which satisfies these conditions is
\begin{equation}
u(s) \simeq {u_0 \over 2}\left( 1- \nvec_1 \cdot \nvec_2\right)
\left[- \left( {1 \over 2} - {l_m \over l} \right) -s  \right]
\label{motor_fil_step}\;,
\end{equation}
where $u_0$ is the (constant) single-motor 'stepping rate'
(step-size/cycle-time).

Assuming $u_0 \gg \gamma$ and substituting
Eq.~(\ref{motor_fil_step}) into Eqs. ~(\ref{vrel_fil_app}),
(\ref{Vnet_fil_app}) and (\ref{omega_micro_fil}), we obtain an
expression for relative velocity of the two thin filaments
connected by a thick filament,
\begin{equation}
{\bf v} \simeq u_0(1-\nvec_1\cdot\nvec_2)\Big[\frac{\bxi}{l}
-\frac{1}{2} \frac{(l-l_m)}{l}(\nvec_2-\nvec_1)\Big]\;,
\label{vrel_contract}
\end{equation}
where ${\bm \xi}= s_1 \nvec_1 - s_2 \nvec_2 + l_m \nvec_m$.  In
obtaining Eq.~ (\ref{vrel_contract}) we used approximated $\nvec_m
\simeq 1/2(\nvec_2 - \nvec_1)$, which holds when the filaments are
antiparallel. By comparing Eqs.~(\ref{vrel_contract}) and
(\ref{omega_micro_fil}) to the general expressions given in Sec.
\ref{sec:Model} we obtain the following estimates
\begin{eqnarray}
&&\talpha_0=-\talpha_1\simeq u_0\;,\\
&& \beta_0=-\beta_1\simeq-\Big(1-\frac{l_m}{l}\Big)u_0\;,\\
&& \gamma_P = \gamma \simeq \kappa/\zeta_r\;.
\end{eqnarray}
The crucial difference between the effect of the stalling
crosslinker considered in Sec.~\ref{sub:stalling} and the
contractile minifilament considered here is of course that in the
present case the rates $\talpha_0$ ($\beta_0$) and $\talpha_1$
($\beta_1$) have opposite signs. This will have important effects
on the system's rheology.

\section{Moment expansion}
\label{app:expansion}
To define the exact moment expansion of the
filament concentration $c(\bfr,\nvec,t)$ we introduce a  set of
irreducible tensors $T^{m}_{i_1i_2...i_m}$ which are equivalent to
the spherical harmonics, but are expressed in Cartesian
coordinates. The components of $T^{m}_{i_1i_2...i_m}$ are
homogeneous polynomials of degree $m$ in the components of the
unit vector $\nvec$, with the properties that they are fully
symmetric in the subscripts $i_1$, $i_2$, ..., $i_m$, and that no
nonvanishing tensor of lower order can be formed by contraction.

Here we denote by $d$ the dimensionality and write the general
expression of the $T^{m}$ for $d=2,3$. Each tensor is orthogonal
to all the other ones and normalized according to a product
defined by
\begin{equation}
\label{product}
\Big(a,b\Big)=\int\frac{d\nvec}{\Omega_d}~a(\nvec)~b(\nvec)\;,
\end{equation}
where $\Omega_d$ is the solid angle in $d$ dimensions, with
$\Omega_2=2\pi$ and $\Omega_3=4\pi$. The first few irreducible
tensors are
\begin{equation}
T^{0}=1\;,
\end{equation}
\begin{equation}
T^1_i=\nhat_i\;,
\end{equation}
\begin{equation}
T^2_{ij}=\nhat_i\nhat_j-\frac{1}{d}\delta_{ij}\;,
\end{equation}
\begin{equation}
T^3_{ijk}=\nhat_i\nhat_j\nhat_k-\frac{1}{d+2}\big[\delta_{ij}\nhat_k+\delta_{ik}\nhat_j+\delta_{jk}\nhat_i\big]\;,
\end{equation}
\begin{equation}
T^4_{ijkl}=\nhat_i\nhat_j\nhat_k\nhat_l
-\frac{1}{d(d+2)}\big[\delta_{ij}\delta_{kl}+\delta_{ik}\delta_{jl}+\delta_{il}\delta_{jk}\big]\;,
\end{equation}
\begin{eqnarray}
T^5_{ijkls}=\nhat_i\nhat_j\nhat_k\nhat_l\nhat_s-
&&\frac{1}{(d+2)(d+4)}\Big[\nhat_i\Delta_{jkls}
  + \nhat_j\Delta_{ikls}\nonumber\\
    &&\nhat_k\Delta_{ijls}+ \nhat_l\Delta_{ijks}
    +\nhat_s\Delta_{ijkl}\Big]\;,
\end{eqnarray}
where repeated indices are summed over and the tensor
$\Delta_{ijkl}$ is given by
\begin{equation}
\label{Delta} \Delta_{ijkl}=\delta_{ij}\delta_{kl}+\delta_{ik}\delta_{jl}+\delta_{il}\delta_{jk}\;,
\end{equation}
it also satisfies $\Delta_{iikk}=d(d+2)$.

The filament concentration has an exact expansion on the basis of
these irreducible tensors, given by
\begin{equation}
c(\bfr,\nvec,t)=\sum_{m=0}^\infty
a^{m}_{i_1i_2...i_m}(\bfr,t)T^{m}_{i_1i_2...i_m}(\nvec)\;,
\end{equation}
where the $m$-th order moment $a^m_{i_1i_2...i_m}$ is a tensor
determined by
\begin{equation}
a^m_{j_1j_2...j_m}(\bfr,t)\int\frac{d\nvec}{\Omega_d}~T^{m}_{j_1j_2...j_m}(\nvec)T^{m}_{i_1i_2...i_m}(\nvec)=\int\frac{d\nvec}{\Omega_d}~T^{m}_{i_1i_2...i_m}(\nvec)c(\bfr,\nvec,t)\;.
\end{equation}
The first three moments are given by
\begin{eqnarray}
&&
a^0=\int\frac{d\nvec}{\Omega_d}~c(\bfr,\nvec,t)=\frac{1}{\Omega_d}\rho(\bfr,t)\;,\\
&&
a^1_i=d\int\frac{d\nvec}{\Omega_d}~\nhat_i~c(\bfr,\nvec,t)=\frac{d}{\Omega_d}\rho(\bfr,t)p_i(\bfr,t)\;,\\
&&
a^2_{ij}=\frac{d(d+2)}{2}\int\frac{d\nvec}{\Omega_d}~\Big(\nhat_i\nhat_j-\frac{1}{d}\delta_{ij}\Big)c(\bfr,\nvec,t)
=\frac{d(d+2)}{2\Omega_d}\rho(\bfr,t)S_{ij}(\bfr,t)\;.
\end{eqnarray}
Here $\rho$, ${\bf p}$ and $S_{ij}$ are the density, polarization,
and nematic order parameter of the rods, respectively. Retaining
only moments up to the third one,  the filament concentration can
be written as
\begin{equation}
\label{c_3moments}
c(\bfr,\nvec,t)\approx\frac{1}{2^{d-1}\pi}\rho(\bfr,t)\Big[1+d\nvec\cdot{\bf
p}(\bfr,t)+\frac{d(d+2)}{2}\hat{Q}_{ij}(\nvec)S_{ij}(\bfr,t)\Big]\;,
\end{equation}
with $\hat{Q}_{ij}=\nhat_i\nhat_j-\frac{1}{d}\delta_{ij}$.

\section{Derivation of coarse-grained currents}
\label{app:currents} In this Appendix we outline the derivation of
the currents and source terms entering the equations for the
density, polarization and alignment tensor and give their general
form. For simplicity we restrict ourselves to the case of long
thin rods, where $D_\parallel=2D_\perp\equiv D$ and $\sigma=1/3$.

It is convenient to separate the translational and rotational
currents  defined in Eqs.~(\ref{Jr}-\ref{JQ}) and
(\ref{RT}-\ref{RQ}) in diffusive, excluded volume and active
contributions,
\begin{eqnarray}
\label{rho_curr_threeterms} &&J_{i}=J^D_{i}+J_i^{\rm ex}+J^A_{i}\;,\nonumber\\
\label{p_curr_threeterms} &&J_{ij}=J^D_{ij}+J_{ij}^{\rm ex}+J^A_{ij}\;,\nonumber\\
\label{Q_curr_threeterms} &&J_{ijk}=J^D_{ijk}+J_{ijk}^{\rm
ex}+J^A_{ijk}\;,
\end{eqnarray}
and
\begin{eqnarray}
\label{p_rot_threeterms} &&R_i=R^D_i+R_i^{\rm ex}+R^{\rm A}_i\;,\\
\label{Q_rot_threeterms} &&R_{ij}=R^D_{ij}+R_{ij}^{\rm ex}+R^{\rm
A}_{ij}\;,
\end{eqnarray}
where each contribution arises from the corresponding term in
Eqs.~(\ref{current}-\ref{current2}).

The diffusive contributions  are evaluated by inserting the
truncated moment expansion for the filament concentration in the
corresponding contributions to the translational and rotational
diffusion currents  in Eqs.~(\ref{current}-\ref{current2}) and performing the
angular average, with the result
\begin{eqnarray}
\label{JDrho} &&{\bf J}^D=\partial_j\sigma_{ij}^D\;,\hspace{0.4in}
\sigma_{ij}^D=-\frac{D}{2}\Big(\frac{3}{2}\delta_{ij}\rho+Q_{ij}\Big)\;,\\
&&\label{p_currD} J^D_{ij}=
-\frac{D}{4}\Big(\delta_{ij}\bnabla\cdot\bfT+\frac{5}{2}\partial_jT_i\Big)\;,\\
&&\label{Q_currD} J^D_{ijk}
=-\frac{D}{16}\Big(\delta_{jk}\partial_i\rho+\delta_{ik}\partial_j\rho-\delta_{ij}\partial_k\rho\Big)
-\frac{D}{6}\Big[\delta_{ik}\partial_lQ_{jl}+\delta_{jk}\partial_l Q_{il}+\frac{7}{2}\partial_k Q_{ij}
-\delta_{ij}\partial_l Q_{kl}\Big]\;,
\end{eqnarray}
and
\begin{eqnarray}
&&\label{p_rotD}
R^D_i=D_rT_i\;,\\
&&\label{Q_rotD} R_{ij}^D=4D_rQ_{ij}\;.
\end{eqnarray}

To evaluate the excluded volume contributions we expand the
concentration in Eq.~(\ref{Vex}) near its value at ${\bf r}_1$ as
in Eq.~(\ref{c_exp}), truncate the moment expansion of the
concentration to third order and perform the angular integrations.
Retaining  terms up to first order in the gradients of the fields
in the currents and up to second order in the source terms, we
obtain
\begin{equation}
\label{Jex_rho}
J_i^{\rm{ex}}=Dv_0\partial_j\Big[-\frac{1}{2}\rho\Big(Q_{ij}+\frac{3}{4}\delta_{ij}\rho\Big)
+\frac{2}{9}Q_{ik}Q_{jk}+\frac{7}{18}\delta_{ij}Q_{kl}Q_{kl}\Big]+\frac{2}{3}Dv_0\rho\partial_j
Q_{ij}\;,
\end{equation}
\begin{eqnarray}
\label{Jex_P}
J_{ij}^{\rm{ex}}&=&-\frac{1}{2}Dv_0\Big[\frac{1}{4}(\Delta_{ijkl}+4\delta_{il}\delta_{jk})T_k\partial_l\rho
-\frac{1}{9}(\Delta_{ijln}+6\delta_{in}\delta_{jl})T_k\partial_n Q_{kl}\nonumber\\
&&-\frac{1}{9}(\Delta_{ijln}-\delta_{ij}\delta_{ln})T_n\partial_k
Q_{kl}-\frac{1}{9}T_k\partial_kQ_{ij}\Big]\;,
\end{eqnarray}
\begin{eqnarray}
\label{Jex_Q} J_{ijk}^{\rm{ex}}&=&\frac{1}{12}Dv_0\Big\{
-\frac{3}{8}(\delta_{ik}\delta_{jl}+\delta_{il}\delta_{jk}-\delta_{ij}\delta_{kl})\partial_l\rho^2\nonumber\\
&&+\frac{1}{3}\rho\partial_l\Big[7\delta_{kl}Q_{ij}+\delta_{ik}Q_{jl}+\delta_{jl}Q_{ik}
+\delta_{il}Q_{jk}+\delta_{jk}Q_{il}-2\delta_{ij}Q_{kl}\Big]\nonumber\\
&&-\Big[7\delta_{kl}Q_{ij}+\delta_{ik}Q_{jl}+\delta_{jl}Q_{ik}+\delta_{il}Q_{jk}
+\delta_{jk}Q_{il}-2\delta_{ij}Q_{kl}\Big]\partial_l\rho\nonumber\\
&&+\frac{1}{3}\partial_l\Big[Q_{ij}Q_{kl}+Q_{ik}Q_{jl}+Q_{il}Q_{jk}-\delta_{ij}Q_{kr}Q_{lr}\nonumber\\
&&+(\delta_{il}Q_{jr}+\delta_{jl}Q_{ir}-\delta_{ij}Q_{lr})Q_{kr}
+(\delta_{ik}Q_{jr}+\delta_{jk}Q_{ir}-\delta_{ij}Q_{kr})Q_{lr}\nonumber\\
&&+\frac{1}{4}(\delta_{ik}\delta_{jl}+\delta_{il}\delta_{jk}-19\delta_{ij}\delta_{kl})Q_{rs}Q_{rs}
+9\delta_{kl}Q_{ir}Q_{jr}\Big]\Big\}\;,
\end{eqnarray}
where $v_0=\frac{2}{\pi}$, and
\begin{eqnarray}
\label{Jex_P_rot} R_i^{\rm{ex}}&=&-\frac{1}{3}D_rv_0\Big[4T_j
Q_{ij}+\frac{1}{6}T_i\nabla^2\rho-\frac{1}{3}T_j\partial_j\partial_i\rho
+\frac{1}{18}T_j\nabla^2 Q_{ij}\nonumber\\
&&+\frac{1}{9}\Big(T_k\partial_k\partial_j
Q_{ij}+T_j\partial_i\partial_k Q_{jk}-T_i\partial_j\partial_k
Q_{jk}\Big)\Big]\;,
\end{eqnarray}
\begin{eqnarray}
\label{Jex_Q_rot}
R_{ij}^{\rm{ex}}&=&-\frac{4}{3}D_rv_0\rho Q_{ij}-\frac{1}{288}D_rv_0\Big[\rho\Big(\partial_i\partial_j\rho-\frac{1}{2}\delta_{ij}\nabla^2\rho\Big)
+(\delta_{il}Q_{jk}+\delta_{jl}Q_{ik}-\delta_{ij}Q_{kl})\partial_k\partial_l\rho\Big]\;.
\end{eqnarray}
To evaluate the active contributions, we insert the gradient
expansion of the concentration and motor density given in
Eqs.~(\ref{c_exp}) and (\ref{m_exp}) in the Eq.~(\ref{Jm}) for the
motor current density and in Eqs. (\ref{Jtact}) and (\ref{Jract}) 
for the filament currents. The integrals over the
lengths $s_1$ and $s_2$ of the filaments can then be evaluated
explicitly. All terms containing odd powers of components of the
filament center of mass separation $\bm\xi$ vanish when averaged
over the rods' length. To evaluate the angular integrals in the
filamment current densities we also expand the translational and
rates $\talpha(\theta)$ and $\beta(\theta)$, as well as the
excluded volume
$|\nvec_1\times\nvec_2|=\sqrt{1-(\nvec_1\cdot\nvec_2)}$, to first
order in the cosine of the angle between the two filaments,
$\nvec_1\cdot\nvec_2$. With this approximation, the motor-induced
linear and angular velocities ${\bf v}_1={\bf V}+{\bf v}/2$ and
$\bm\omega_1$ are written as
\begin{eqnarray}
{\bf
v}&=&\frac{\beta_0}{2}(\nvec_2-\nvec_1)+\frac{\talpha_0}{2}\bm\xi\nonumber\\
&&+\Big[\frac{\beta_1}{2}(\nvec_2-\nvec_1)+\frac{\talpha_1}{2}\bm\xi\Big](\nvec_1\cdot\nvec_2)+{\cal
O}((\nvec_1\cdot\nvec_2)^2)\;,
\end{eqnarray}
\begin{eqnarray}
{\bf
V}&=&-\frac{\sigma}{4}\beta_0(\nvec_1+\nvec_2)+\frac{\sigma}{4}\talpha_0(\nvec_1s_1+\nvec_2s_2)\nonumber\\
&&-\frac{\sigma}{4}\big[\beta_1-\beta_0(1-\sigma)\big](\nvec_1\cdot\nvec_2)(\nvec_1+\nvec_2)\nonumber\\
&&+\frac{\sigma}{4}\big[\talpha_1(\nvec_1s_1+\nvec_2s_2)-\talpha_0(1-\sigma)
(\nvec_1s_2+\nvec_2s_1)\big](\nvec_1\cdot\nvec_2) +{\cal
O}((\nvec_1\cdot\nvec_2)^2)\;,
\end{eqnarray}
\begin{eqnarray}
\bm\omega_1=2\big[\gamma_P+\gamma_{NP}(\nvec_1\cdot\nvec_2)\big]\nvec_1\times\nvec_2
+{\cal O}((\nvec_1\cdot\nvec_2)^2)\;,
\end{eqnarray}
and $|\nvec_1\times\nvec_2|\approx 1$.  As indicated in the main
text, contributions of higher order in $\nvec_1\cdot\nvec_2$ only
change the values of the numerical coefficients of the various
terms in the expressions for the currents given below, but do not
contribute any qualitatively new terms.

Finally, we insert the moment expansion of the filament
concentration $c(\bfr,\hat{\bm\nu},t)$, truncate it to the first
three moments, as given in Eq.~(\ref{c_3moments}), and evaluate
the active contributions to the various current densities defined
in Eqs.~(\ref{Jr}-\ref{JQ}) and (\ref{RT}-\ref{RQ}).  The
calculation of the angular integrals is quite lengthy and has been
carried out with Maple.

The motor current density is given by
\begin{eqnarray}
\label{Jm_exp}
J^m_i&=&mT_i+\frac{l^3}{48}\Big[T_j\partial_i\partial_jm+\frac{1}{2}T_i\nabla^2m\Big]+{\cal
O}(\nabla^3)\;.
\end{eqnarray}

The active contribution to the current density is naturally
written as the sum of two parts
\begin{eqnarray}
\label{Jrho_twoparts} J^A_{i}({\bf r},t)
=\rho\mathcal{V}_i+\partial_j\sigma^A_{ij}\;,
\end{eqnarray}
where
\begin{eqnarray}
\label{VD}
\rho\mathcal{V}_i&=&-\frac{\tm}{6}\Big(\frac{2\beta_0}{3}+\frac{\beta_1}{2}\Big)\rho
T_i-\frac{\tm}{6}\Big(\beta_1-\frac{2\beta_0}{3}\Big)
Q_{ij}T_j+\frac{\tm\alpha_0}{3}\Big(Q_{ij}\partial_j\rho-\rho\partial_jQ_{ij}\Big)\nonumber\\
&&+\frac{1}{3}\Big[\alpha_0\rho\Big(Q_{ij}+\frac{1}{2}\delta_{ij}\rho\Big)
-\frac{2}{3}\alpha_0\Big(Q_{ik}+\frac{1}{2}\delta_{ik}\rho\Big)\Big(Q_{kj}+\frac{1}{2}\delta_{kj}\rho\Big)\nonumber\\
&&+\frac{\alpha_1}{2}\Big(T_iT_j+\frac{1}{2}\delta_{ij}T^2\Big)\Big]\partial_jm\;,
\end{eqnarray}
and the active contribution to the stress tensor, $\sigma^A_{ij}$,
is given by
\begin{eqnarray}
\label{sigA_ij} \sigma^A_{ij}
=\alpha_0\tm\rho\Big(Q_{ij}+\frac{1}{2}\delta_{ij}\rho\Big)
+\frac{\alpha_1}{2}\tm\Big(T_iT_j+\frac{1}{2}\delta_{ij}T^2\Big)
\;,
\end{eqnarray}
with $\alpha_0=\talpha_0/48$ and $\alpha_1=\talpha_1/48$. The
drift $\mathcal{V}_i$ vanishes in a passive system and arises
entirely from the contribution to the active current from the net
velocity ${\bf V}$ of the pair. It is in fact proportional to
$\sigma=(\zeta_\perp-\zeta_\parallel)/(\zeta_\perp+\zeta_\parallel)$
and vanishes for isotropic objects. The term proportional to
$\alpha_0$ in the stress tensor describes the built-up of density
inhomogeneities via filament bundling and has an effect opposite
to that of conventional diffusion. As shown below, this is the
main term responsible for driving the instability of homogeneous
states.

The active contributions to the translational and rotational
polarization currents are given by
\begin{eqnarray}
\label{p_currA} J^{\rm A}_{ij}&=& \tm\frac{\beta_0}{6}T_iT_j
-\frac{\tm}{6}(\beta_1-\beta_0/6)\Big[T_iT_j+\frac{1}{2}\delta_{ij}T^2\Big]\nonumber\\
&&-\tm\frac{\beta_0}{3}\rho\Big(Q_{ij}+\frac{1}{2}\delta_{ij}\rho\Big)
+\frac{\tm}{6}(\beta_1+\beta_0/3)\Big(Q_{ik}+\frac{1}{2}\delta_{ik}\rho\Big)\Big(Q_{jk}+\frac{1}{2}\delta_{jk}\rho\Big)
\nonumber\\
&&+\frac{\alpha_0}{3}\tm
\Big(T_i\partial_j+T_j\partial_i+\delta_{ij}{\bf
T}\cdot\bm\nabla\Big)\rho
+\frac{\alpha_1}{4}\tm\rho T_{ij}+\frac{\alpha_0}{3}\tm T_j\partial_i\rho\nonumber\\
&&+\frac{2\alpha_0}{3}\tm
T_j\partial_kQ_{ik}+\frac{\alpha_0}{18}\tm\Big[\delta_{ij}T_l\partial_kQ_{kl}-T_k\partial_kQ_{ij}
+T_i\partial_kQ_{kj}-T_k\partial_jQ_{ik}\Big]\nonumber\\
&&+\frac{\alpha_1}{6}\tm Q_{jk}T_{ki}
+\frac{2\alpha_1}{9}\tm\Big[\frac{1}{2}\delta_{ij}Q_{kl}T_{kl}-Q_{ij}\bm\nabla\cdot{\bf
T} +Q_{ik}T_{kj} +Q_{jk}T_{ki}\Big]\nonumber\\
&&+\frac{1}{6}\big(\alpha_1+11\alpha_0/6\big)\rho\Big(T_i\partial_j+T_j\partial_i+\delta_{ij}{\bf
T}\cdot\bm\nabla\Big)\tm\nonumber\\
&&+\frac{2\alpha_1}{9}\Big[
Q_{jk}\big(T_k\partial_i+T_i\partial_k\big)+\delta_{ij}Q_{kl}T_l\partial_k\Big]\tm\nonumber\\
&&+\frac{1}{9}\big(2\alpha_1-\alpha_0/2\big)\Big[
Q_{ik}\big(T_j\partial_k+T_k\partial_j\big)+Q_{ij}T_k\partial_k\Big]\tm
 \;,
\end{eqnarray}
where
\begin{equation}
T_{ij}=\partial_iT_j+\partial_jT_i+\delta_{ij}\bm\nabla\cdot{\bm
T}\;.
\end{equation}
and
\begin{eqnarray}
\label{p_rotA} R^{\rm A}_i&=&-\gamma_P \tm\rho T_i+(2\gamma_P-\gamma_{NP})\tm T_jQ_{ij}\nonumber\\
&&-\frac{\gamma_P}{24}\Big\{\frac{1}{4}\rho(3\delta_{ij}\nabla^2-2\partial_i\partial_j)(\tm T_j)\nonumber\\
&&
-\frac{1}{3}\Big[Q_{ij}(\delta_{jk}\nabla^2+2\partial_j\partial_k)(\tm
T_k)
+Q_{jk}\partial_j\Big(2\partial_i(\tm T_k)-5\partial_k(\tm T_i)\Big)\Big]\nonumber\\
&&+\frac{1}{4}\tm\rho(\delta_{ij}\nabla^2+2\partial_i\partial_j)T_j
-\frac{1}{2}\tm Q_{ij}(\delta_{jk}\nabla^2+2\partial_j\partial_k)T_k\Big\}\nonumber\\
&&-\frac{\gamma_{NP}}{24}\Big\{\frac{1}{3}\Big[T_j(\delta_{ik}\nabla^2-\partial_i\partial_k)(\tm Q_{jk})\nonumber\\
&&+2T_k\partial_k\partial_j(\tm
Q_{ij})-\frac{1}{2}T_i\partial_j\partial_k(\tm Q_{jk})\Big]
+\frac{1}{4}\tm T_j(\partial_j\partial_i-\frac{1}{2}\delta_{ij}\nabla^2)\rho\nonumber\\
&&+\frac{1}{6}\tm\Big[T_j(\delta_{ik}\nabla^2+2\partial_i\partial_k)Q_{jk}+2T_k\partial_k\partial_jQ_{ij}
-2T_i\partial_j\partial_kQ_{jk}\Big]\Big\}\;.
\end{eqnarray}

Finally, the translational and rotational contributions to the
alignment tensor current are
\begin{eqnarray}
\label{Q_currA} J^A_{ijk}&=&
\frac{\tm}{12}\Big\{-\frac{1}{4}\Big(\frac{10}{3}\beta_0+\beta_1\Big)\rho(\delta_{ik}T_j+\delta_{jk}T_i-\delta_{ij}T_k)
+\frac{1}{3}\Big(\frac{19}{3}\beta_0-2\beta_1\Big)T_k Q_{ij}\nonumber\\
&&+\frac{1}{6}\Big(\frac{5}{3}\beta_0-\beta_1\Big)(T_jQ_{ik}+T_iQ_{jk}-\delta_{ij}T_lQ_{kl})\nonumber\\
&&+\frac{1}{3}\Big(\frac{1}{3}\beta_0-2\beta_1\Big)T_l(\delta_{ik}Q_{jl}+\delta_{jk}Q_{il}-\delta_{ij}Q_{kl})\Big\}\nonumber\\
&&+\frac{1}{9}\alpha_0\Big\{\tm\Big[\frac{1}{4}(\delta_{ik}\delta_{jn}+\delta_{jk}\delta_{in}-\delta_{ij}\delta_{kn})\rho
+\frac{19}{3}\delta_{kn}Q_{ij}\nonumber\\
&&+\frac{1}{3}(\delta_{jn}Q_{ik}+\delta_{ik}Q_{nj}+\delta_{in}Q_{jk}+\delta_{jk}Q_{in}-2\delta_{ij}Q_{kn})
\Big]\partial_l(Q_{ln}+\frac{1}{2}\delta_{ln}\rho)\nonumber\\
&&+\frac{11}{2}\Big[\frac{1}{4}(\delta_{ik}\delta_{jl}+\delta_{il}\delta_{jk}-\delta_{ij}\delta_{kl})\rho
+\frac{1}{3}\delta_{kl}Q_{ij}\nonumber\\
&&+\frac{1}{3}(\delta_{jl}Q_{ik}+\delta_{ik}Q_{jl}+\delta_{il}Q_{jk}+\delta_{jk}Q_{il}-2\delta_{ij}Q_{kl})\Big]
\partial_l(\tm\rho)\nonumber\\
&&-\Big[\frac{1}{4}(\delta_{in}\delta_{jl}+\delta_{il}\delta_{jn}-\delta_{ij}\delta_{ln})\rho
+\frac{1}{3}\delta_{ln}Q_{ij}\nonumber\\
&&+\frac{1}{3}(\delta_{il}Q_{jn}+\delta_{jn}Q_{il}+\delta_{jl}Q_{in}+\delta_{in}Q_{jl}-2\delta_{ij}Q_{ln})\Big]
\partial_l(\tm Q_{kn})\Big\}\nonumber\\
&&+\frac{1}{6}\alpha_1\Big\{\frac{1}{4}\tm(\delta_{in}T_j+\delta_{jn}T_i-\delta_{ij}T_n)
\partial_l(\delta_{kl}T_n+\delta_{kn}T_l+\delta_{ln}T_k)\nonumber\\
&&+\frac{1}{3}(\Delta_{vijkln}-3\delta_{ij}\Delta_{klnv})T_v\partial_l(\tm T_n)\Big\}\;,
\end{eqnarray}
where
\begin{equation}
\Delta_{ijkl}=\delta_{ij}\delta_{kl}+\delta_{ik}\delta_{jl}+\delta_{il}\delta_{jk}\;.
\end{equation}
\begin{eqnarray}
\Delta_{ijklnp}=\delta_{ij}\Delta_{klnp}+\delta_{ik}\Delta_{jlnp}+\delta_{il}\Delta_{kjnp}
+\delta_{in}\Delta_{kljp}+\delta_{ip}\Delta_{klnj}\;.
\end{eqnarray}
and
\begin{eqnarray}
R_{ij}^A&=&-2\gamma_P\tm(T_iT_j-\frac{1}{2}\delta_{ij}T^2)
-\gamma_{NP}\tm\rho Q_{ij}\nonumber\\
&&-\frac{\gamma_P}{48}\Big\{\frac{\tm}{2}(T_i\nabla^2T_j+T_j\nabla^2T_i)
+\tm(T_i\partial_j+T_j\partial_i)\partial_kT_k
-\frac{1}{2}\delta_{ij}\tm(T_k\nabla^2T_k+2T_k\partial_k\partial_l T_l)\nonumber\\
&&+\frac{2}{3}\Big[T_i\nabla^2(\tm T_j)+T_j\nabla^2(\tm T_i)
-(T_i\partial_j+T_j\partial_i)\partial_k(\tm T_k)-T_k\partial_i\partial_j(\tm T_k)\nonumber\\
&&-\frac{1}{2}\delta_{ij}\Big(T_k\nabla^2(\tm
T_k)+2T_k\partial_k\partial_l(\tm T_l)\Big)
+2T_k\partial_k\Big(\partial_i(\tm T_j)+\partial_j(\tm T_i)\Big)\Big]\Big\}\nonumber\\
&&-\frac{\gamma_{NP}}{48}\Big\{\frac{\tm}{2}\rho\Big[\partial_i\partial_j\rho-\frac{1}{2}\delta_{ij}\nabla^2\rho
+\frac{2}{3}(\nabla^2Q_{ij}+2\partial_i\partial_kQ_{jk}+2\partial_j\partial_kQ_{ik}-2\delta_{ij}\partial_k\partial_lQ_{kl})\Big]\nonumber\\
&&+\frac{\tm}{3}\Big[Q_{ik}\partial_j\partial_k\rho+Q_{jk}\partial_i\partial_k\rho-Q_{ij}\nabla^2\rho
-\delta_{ij}Q_{kl}\partial_k\partial_l\rho\Big]\nonumber\\
&&+\frac{2\tm}{9}\Big[Q_{ik}\nabla^2Q_{jk}+Q_{jk}\nabla^2Q_{ik}+2Q_{il}(\partial_j\partial_kQ_{kl}+\partial_k\partial_lQ_{jk})\nonumber\\
&&+2Q_{jl}(\partial_i\partial_k Q_{kl}+\partial_k\partial_l
Q_{ik})-4Q_{ij}\partial_k\partial_l Q_{kl}
-\delta_{ij}Q_{kl}(\nabla^2Q_{kl}+4\partial_l\partial_r Q_{kr})\Big]\nonumber\\
&&+\frac{1}{3}\rho\Big[2\nabla^2(\tm
Q_{ij})+\partial_i\partial_k(\tm Q_{jk})+\partial_j\partial_k(\tm
Q_{ik})
-\delta_{ij}\partial_k\partial_l(\tm Q_{kl})\Big]\nonumber\\
&&+\frac{1}{3}\Big[Q_{ik}\nabla^2(\tm Q_{jk})+Q_{jk}\nabla^2(\tm
Q_{ik})
-2Q_{il}\Big(\partial_j\partial_k(\tm Q_{kl})-\partial_k\partial_l(\tm Q_{jk})\Big)\nonumber\\
&&-2Q_{jl}\Big(\partial_i\partial_k(\tm
Q_{kl})-\partial_l\partial_k(\tm Q_{ik})\Big)
+2Q_{kl}\Big(\partial_i\partial_k(\tm Q_{jl})+\partial_j\partial_k(\tm Q_{il})\Big)\nonumber\\
&&+3Q_{kl}\partial_k\partial_l(\tm Q_{ij})
-Q_{ij}\partial_k\partial_l(\tm Q_{kl})-Q_{kl}\partial_i\partial_j(\tm Q_{kl})\nonumber\\
&&-\delta_{ij}Q_{kl}\Big(\frac{1}{2}\nabla^2(\tm
Q_{kl})+2\partial_l\partial_r(\tm Q_{kr})\Big)\Big]\Big\}\;.
\end{eqnarray}

 The general
nonlinear equations are fairly complicated, but the various terms
have simple physical interpretations, as will become apparent
below. The terms proportional to $\alpha_0$ and $\alpha_1$ tend to
bundle filaments together, therefore enhancing density
fluctuations. The terms proportional to $\beta_0$ and $\beta_1$
tend to align the filaments in the direction of the polarization
and thus suppress polarization fluctuations.
The $\gamma_P$ and $\gamma_{NP}$ terms rotate and align filament and
play a crucial role in controlling the possible homogeneous states
of the system.

\section{Role of higher order gradients}
\label{app:fourth_order}

\begin{figure}
\center \resizebox{0.7\textwidth}{!}{%
  \includegraphics{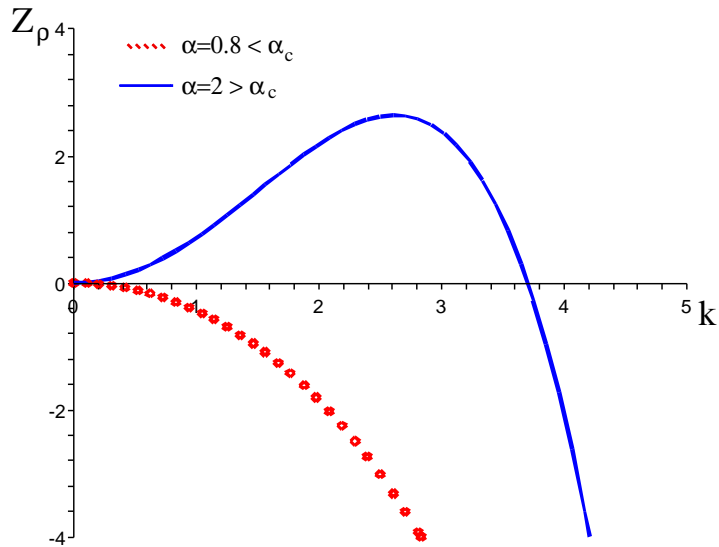}
}

\caption{(color online) The density mode $z_\rho$ as a function of $k$ for
different values of the parameter $\alpha$ for $\tm=1$,
$\rho_0=1$, $\alpha^{'}_c \simeq 0.57$ and $\alpha_c \simeq 1.23$.
For $\alpha^{'}_c <\alpha <\alpha_c$ the isotropic state is always
stable, and for $\alpha>\alpha_c$ the isotropic state is unstable
for long wavelength.}
\label{regions_4th}       
\end{figure}

Here we discuss the role of terms of order higher than second in
the gradients of the hydrodynamic fields in controlling the
bundling instability. For simplicity we only consider the
instability of the isotropic state. In this case the only
hydrodynamic variable is the filament density. Including terms of
order up to $k^4$, the dynamics of the Fourier components of
density fluctuations defined in Eq.~(\ref{fourier_rho}) is
governed by
\begin{eqnarray}
\label{density4th} \partial_t\rho_{\bf
k}=-k^2\Big[\frac{3}{4}D(1+v_0\rho_0)-\alpha m\rho_0\Big]\rho_{\bf
k} +\frac{1}{96}\Big[\frac{13}{4}Dv_0\rho_0-\frac{19}{5}\alpha
m\rho_0\Big]k^4\rho_{\bfk}\;.
\end{eqnarray}
Their decay is controlled by a single diffusive mode, given by
\begin{equation}
\label{densitymode4th} z_\rho=-C_2k^2+C_4k^4\;, 
\end{equation}
where
\begin{eqnarray}
&&C_2=m\rho_0(\alpha_c-\alpha)\;,\\
&& C_4=\frac{19}{480}m\rho_0(\alpha'_c-\alpha)\;,
\end{eqnarray}
and
\begin{eqnarray}
&&\alpha_c=\frac{3Dv_0}{4m}\frac{1+v_0\rho_0}{v_0\rho_0}\;,\\
&&\alpha'_c=\frac{65Dv_0}{72m}\;.
\end{eqnarray}
At the low filament densities where the isotropic phase exists the
value $\alpha_c$ where the coefficient $C_2$ changes sign grows
rapidly with filament density, while at the value $\alpha'_c$
where the coefficient $C_4$ changes sign is independent of
$\rho_0$. We therefore expect $\alpha_c>\alpha'_c$ in the region
of interest. We can then identify three regions:
\begin{itemize}
\item For $\alpha<\alpha'_c$ both $C_2$ and $C_4$ are positive.
Long wavelength density fluctuations always decay and the
isotropic state is stable. The growth rate defined in
Eq.~(\ref{densitymode4th}) becomes positive for $k>k_0$, with
$k_0=\sqrt{C_2/C_4}$, but this short scale instability is outside
the range of validity of the present work. We expect that it will
be suppressed by terms of even higher order in the gradients.
\item For $\alpha'_c<\alpha<\alpha_c$ we have $C_2>0$ and $C_4<0$
and the isotropic state is always stable. \item For
$\alpha>\alpha_c$ the eigenvalue $z_\rho(k)$ controlling the
dynamics of density fluctuations becomes positive for $k<k_0$. In
this regime long wavelength density fluctuations grow in time and
the isotropic state is unstable. The isotropic state is stabilized
again at short scales, $k>k_0=\sqrt{C_2/C_4}$.
\end{itemize}
The location of the instability in the $(\alpha,\rho_0)$ is not
affected by terms beyond quadratic in the gradients. These terms
do, however, introduce a length scale corresponding to the
wavevector
$k_0=\sqrt{C_2/C_4}\sim\sqrt{(\alpha-\alpha_c)/(\alpha-\alpha'_c)}$
beyond which the isotropic state is stabilized by short scale
effects as seen in Fig.~\ref{regions_4th}. The wavevector of the
fastest growing mode in the unstable region is
$k_m=\sqrt{C_2/2C_4}\sim\sqrt{\alpha_c/(\alpha_c-\alpha'_c)}\sim\epsilon^{1/2}$,
that vanishes with the distance
$\epsilon=(\alpha-\alpha_c)/\alpha_c$ from the instability.

\bibliography{references.bib}

\begin{thebibliography}{42}
\bibitem[{tor()}]{torsional_stiffness}
\bibinfo{note}{In general the torsional stiffness can also vary along the
  filament, i.e., $\kappa=\kappa(s)$}.
\bibitem[{non()}]{nonlinear_polarized}
\bibinfo{note}{Spatial inhomogeneities in the torsional spring constant of the
  crosslinking clusters yield additional terms in Eq.~(\ref{phateqn_pol_state})
  proportional to $d\gamma_P(s)/ds$ . These terms have the same structure as
  those proportional to $\beta$ and therefore do not change the form of the
  polarization equation. There is, however, an important difference between the
  contributions proportional to $\beta$ and those proportional to
  $d\gamma_P(s)/ds$, in that while the former always arise from motor activity,
  the latter could also be originated by stationary crosslinks}.
\end{thebibliography}

\end{document}